\documentclass[sigconf,balance=false,authorversion]{acmart}
\usepackage{popets}
\usepackage[flushleft]{threeparttable}
\usepackage{amsthm}
\usepackage{booktabs}
\usepackage{lscape}
\usepackage{enumitem}
\usepackage{graphicx}
\usepackage{adjustbox}
\usepackage{longtable}
\usepackage[normalem]{ulem}
\useunder{\uline}{\ul}{}
\usepackage{ wasysym }
\usepackage{multirow}

\setcopyright{popets}
\copyrightyear{2024}

\acmYear{2024}
\acmVolume{2024}
\acmNumber{3}
\acmDOI{10.56553/popets-2024-0079}
\acmISBN{}
\acmConference{Proceedings on Privacy Enhancing Technologies}
\usepackage{tikz}
\newcommand*\emptycirc[1][1ex]{\tikz\draw (0,0) circle (#1);} 
\newcommand*\halfcirc[1][1ex]{\begin{tikzpicture}
  \draw[fill] (0,0)-- (90:#1) arc (90:270:#1) -- cycle ;
  \draw (0,0) circle (#1);
  \end{tikzpicture}}
\newcommand*\fullcirc[1][1ex]{\tikz\fill (0,0) circle (#1);} 

\newcommand*\exemptycirc[1][1ex]{\scalebox{1.5}{$\otimes$}} 

\newcommand{\shiftleft}[2]{\makebox[0pt][r]{\makebox[#1][l]{#2}}}
\NewDocumentCommand{\rot}{O{35} O{1em} m}{\makebox[#2][l]{\rotatebox{#1}{#3}}}
\NewDocumentCommand{\rotDown}{O{35} O{1em} m}{
    \makebox[#2][l]{
        \shiftleft{10pt}{
            \raisebox{-0.4cm }[2.5cm][0pt] {
                \rotatebox{#1}{
                    {#3}
                }
            }
        }
    }
}
\NewDocumentCommand{\down}{O{35} O{1em} m}{
    \makebox[#2][l]{
        \shiftleft{10pt}{
            \raisebox{-0.4cm }[0cm][0pt] {
                {#3}    
            }
        }
    }
}

\theoremstyle{definition}
\newtheorem{threat}{\indent Threat}[section]

\NewDocumentCommand{\threatcmd}{m m m m m}{\noindent\begin{threat} \label{#1}
        #2 \\
        \noindent\textbf{Risk:} #3 \\
        \noindent\textbf{Constraint:} #4 \\
        \noindent\textbf{Mitigations:} #5 \\
    \end{threat}
}

\graphicspath{ {./pets24_3/img/} }

\begin{document}
\title[SoK: Trusting Self-Sovereign Identity]{SoK: Trusting Self-Sovereign Identity}

\author{Evan Krul}
\orcid{0009-0007-3190-0238}
\affiliation{\institution{University of New South Wales}
  \institution{Cyber Security Cooperative Research Centre}
  \city{Sydney}
  \state{NSW}
  \country{Australia}}
\email{e.krul@unsw.edu.au}

\author{Hye-young Paik}
\affiliation{\institution{University of New South Wales}
  \city{Sydney}
  \state{NSW}
  \country{Australia}}
\email{h.paik@unsw.edu.au}

\author{Sushmita Ruj}
\affiliation{\institution{University of New South Wales}
  \city{Sydney}
  \state{NSW}
  \country{Australia}}
\email{sushmita.ruj@unsw.edu.au}

\author{Salil S. Kanhere}
\affiliation{\institution{University of New South Wales}
  \city{Sydney}
  \state{NSW}
  \country{Australia}}
\email{salil.kanhere@unsw.edu.au}

\renewcommand{\shortauthors}{Krul et al.}

\begin{abstract}
    Digital identity is evolving from centralized systems to a decentralized approach known as Self-Sovereign Identity (SSI). SSI empowers individuals to control their digital identities, eliminating reliance on third-party data custodians and reducing the risk of data breaches. However, the concept of trust in SSI remains complex and fragmented. This paper systematically analyzes trust in SSI in light of its components and threats posed by various actors in the system. As a result, we derive three distinct trust models that capture the threats and mitigations identified across SSI literature and implementations. Our work provides a foundational framework for future SSI research and development, including a comprehensive catalogue of SSI components and design requirements for trust, shortcomings in existing SSI systems and areas for further exploration.
 \end{abstract}


\maketitle

\section{Introduction} \label{introduction}
Digital identity encompasses not only usernames and passwords but also our online presentation, interactions with others, and how the world perceives us~\cite{allenPathSelfSovereignIdentity2016}. In 2023, Louisiana passed a law requiring all websites containing adult content to verify the user's age. Louisiana's law mandated users present their state-issued digital identities, such as driver's licenses, to access these websites~\cite{culeLouisianaAgeVerification2023}. This is not a singular event; digital identities are becoming ubiquitous tools in many applications, including age verification, know-your-customer compliance~\cite{schlattDesigningFrameworkDigital2021, moyanoKYCOptimizationUsing2017}, and academic credential verification~\cite{wolzDigitalCredentialsHigher2021}. Digital identity has expanded beyond mere name and email addresses in an online account.

In the current ecosystem, digital identity is managed centrally through \textit{siloed} (one identity per service) or \textit{federated} (single sign-on) identity systems~\cite{shimFederatedIdentityManagement2005}. These centralized identity management systems typically rely on a single entity, such as a government agency or a private entity, to issue and manage identities. This concentration of power has created a single point of failure, making these systems susceptible to data breaches and misuse. In recent years, there have been numerous high-profile breaches that highlight the inherent risks of centralized identity management~\cite{cadwalladrRevealed50Million, UnauthorizedAccessOkta2023, EquifaxDataBreach, optusOptusNotifiesCustomers}. 

Furthermore, the reliance on centralized authorities has placed undue trust on these data custodians. Users are forced to relinquish control over their personal identity information, with limited transparency and accountability regarding how their data is used. This lack of control raises concerns about potential privacy violations, misuse of identity data, and the possibility of the central authority gaining access to sensitive information about the usage of an identity~\cite{soltaniSurveySelfSovereignIdentity2021, hinchliffeWEFPushesDigital2022}. 

In response to these concerns, a new identity paradigm has emerged: Self-Sovereign Identity (SSI), aiming to revolutionize identity management by empowering individuals to take control of their digital identity in the absence of a centralized authority~\cite{allenPathSelfSovereignIdentity2016}. The essence of SSI lies in the issuance and verification of user credentials. These credentials, issued by trusted entities, attest to the authenticity of a particular attribute of a user's identity. By holding the credentials, users eliminate reliance on third-party data custodians, reduce the potential for data breaches, and ensure they retain control over their identity. 

While SSI bears the same underlying principle of fostering trust between unknown parties as previous systems~\cite{cameronLawsIdentity}, the extent of what ``trust'' means in this context has undergone considerable discussion. In both conceptual and implementation level discussions of SSI, there are disagreements on what trust is already available between participants and what trust needs to be built up through the design of the system~\cite{arnoldZeroKnowledgeProofsNot2019b, hardmanNoParadoxHere}. The discussions and summaries in this paper will illustrate how complex and fragmented the field of SSI has become. Without a clear agreement on the threats the participants face, there can be no guidance on what SSI must address in its design and requirements. 

The paper aims to systematically analyze the elements of SSI in light of trust requirements and assumptions. We do this by breaking down SSI into its individual components to understand the underlying trust assumptions (or alternatively threats) that motivated their design requirements. This analysis provides insights into the design of an SSI system and identifies areas for further research and development. A detailed description of the data collection and analysis process is included in Appendix~\ref{sec:app:methodology}.

\clearpage
\subsection{Contributions}
We claim the following three major contributions:

\begin{enumerate}
    \item A comprehensive analysis of trust in SSI, introducing three distinct trust models to capture the varied levels of assumed trust across literature and implementations. These models are built by identifying the components of SSI and the threats they face from various actors in the system.
    \item An analysis of known implementations of SSI, both in academic literature and industry products, detailing their trust assumptions made and the implementation approaches taken to address the assumed threats identified in (1). 
    \item A foundational framework for future SSI research and development by providing a comprehensive catalogue of SSI components and design requirements according to trust models. We identify shortcomings in existing SSI systems and highlight areas for further exploration.
\end{enumerate}

The remainder of this paper is structured as follows: Section \ref{sec:related_work} provides an overview of related work. Section \ref{sec:2_1_ssi_background} introduces the core concepts of SSI, laying the foundation for subsequent discussions. Section \ref{sec:2_toip_background} uses these core concepts to establish a minimal trust model for SSI, highlighting the design requirements necessary to mitigate associated threats. Section \ref{sec:3_extensions} delves into the threats that issuers, identity owners, and service providers face, extending beyond the minimal trust model. Section \ref{sec:trust_models} consolidates these threats into two additional threat models. Section~\ref{sec:ssi_state} evaluates the state-of-the-art academic literature and industry implementations, and Section~\ref{sec:open_work} concludes with a discussion of open research directions.

\subsection{Related Work} \label{sec:related_work}

We situate our work within the landscape of well-established surveys on SSI and closely related topics. Through six categorizations, we differentiate our contribution from existing literature and present a summary in Table~\ref{tab:related_work}.

\textbf{Academic Literature and Industry Implementations; }
We present a comprehensive analysis of both academic literature and industry implementations. There have been several surveys that collect and analyze academic literature on SSI. \citet{soltaniSurveySelfSovereignIdentity2021} performed a thorough survey and provided a detailed and in-depth survey of SSI. They provided compelling motivation for SSI and then introduced core components with a strong focus on academic literature, briefly considering a subset of industry implementations. \citet{schardongSelfSovereignIdentitySystematic2022a} extensively covered publishing venues and provided a taxonomy of academic literature. \citet{zhuIdentityManagementSystems2018} described a sub-set of academic and industry contributions focused on blockchain identity for the internet-of-things. \citet{muhleSurveyEssentialComponents2018} did not consider any industry contributions, only surveying academic contributions. \citet{ernstbergerSoKDataSovereignty} and \citet{bistarelliSurveyDecentralizedIdentifier} both broadly collected DID methods from academia and industry. \citet{limTrustModelsBlockchainBased2022} and \citet{satybaldySelfSovereignIdentitySystems2020} evaluated a sub-set of industry blockchain identity systems. \citet{grunerAnalyzingComparingSecurity2023a} only evaluated a single industry implementation.

\textbf{Catalogue of Components;}
This work presents a catalogue of components of SSI and their design requirements.~\citet{schardongSelfSovereignIdentitySystematic2022a} independently collected a large catalogue, while~\citet{ernstbergerSoKDataSovereignty} presented a limited catalogue.

\textbf{Threats and Trust Assumptions;}
A key focus of our contribution is trust in SSI and the associated threats.~\citet{AriesrfcsConcepts0207credentialfraudthreatmodel} presented the most complete SSI threat model focused on preventing credential fraud.~\citet{grunerAnalyzingComparingSecurity2023a} provided a threat model for a single implementation, while~\citet{ernstbergerSoKDataSovereignty} had limited discussion on threats faced.~\citet{limTrustModelsBlockchainBased2022} introduced a trust model for blockchain implementations.

\textbf{Foundation Framework;}
\citet{satybaldySelfSovereignIdentitySystems2020} is the only contribution presenting a framework to evaluate future work. They evaluated five SSI constructions with a framework inspired by \cite{allenPathSelfSovereignIdentity2016}.

\textbf{Not Limited to Blockchain Identity;}
A key distinction with our contribution is that we do not present a survey focused on blockchain-based SSI contributions. 
\citet{schardongSelfSovereignIdentitySystematic2022a} were the only other work not to have a scope limited to blockchain identities.

\begin{table}[h]
\vspace{-1em}
\caption{Related Works Comparison}
\label{tab:related_work}
\begin{adjustbox}{width=0.47\textwidth,center}
\begin{tabular}{@{}llllllllllllp{1.75cm}@{}}
\toprule
                                        & \textbf{\rot{Our Work}} & \textbf{\rot{\citet{grunerAnalyzingComparingSecurity2023a}}} & \textbf{\rot{\citet{soltaniSurveySelfSovereignIdentity2021}}} & \textbf{\rot{\citet{satybaldySelfSovereignIdentitySystems2020}}} & \textbf{\rot{\citet{zhuIdentityManagementSystems2018}}} & \textbf{\rot{\citet{muhleSurveyEssentialComponents2018}}} & \textbf{\rot{\citet{schardongSelfSovereignIdentitySystematic2022a}}} & \textbf{\rot{\citet{limTrustModelsBlockchainBased2022}}} & \textbf{\rot{\citet{ernstbergerSoKDataSovereignty}}} & \textbf{\rot{\citet{AriesrfcsConcepts0207credentialfraudthreatmodel}}} & \textbf{\rot{\citet{bistarelliSurveyDecentralizedIdentifier}}} & \textbf{} \\ \midrule
\textbf{Academic Literature}            & \fullcirc               & \emptycirc                                                   & \fullcirc                                                     & \emptycirc                                                       & \halfcirc                                               & \halfcirc                                                 & \fullcirc                                                            & \emptycirc                                               & \halfcirc                                            & \emptycirc                                                             & \halfcirc                                                      &           \\
\textbf{Industry Implementations}       & \fullcirc               & \halfcirc                                                    & \halfcirc                                                     & \halfcirc                                                        & \halfcirc                                               & \emptycirc                                                & \emptycirc                                                           & \halfcirc                                                & \fullcirc                                            & \emptycirc                                                             & \halfcirc                                                      &           \\
\textbf{Catalogue of Components}        & \fullcirc               & \emptycirc                                                   & \emptycirc                                                    & \emptycirc                                                       & \emptycirc                                              & \emptycirc                                                & \fullcirc                                                            & \emptycirc                                               & \halfcirc                                            & \emptycirc                                                             & \emptycirc                                                     &           \\
\textbf{Threats and Trust Assumptions}  & \fullcirc               & \halfcirc                                                    & \emptycirc                                                    & \emptycirc                                                       & \emptycirc                                              & \emptycirc                                                & \emptycirc                                                           & \halfcirc                                                & \halfcirc                                            & \fullcirc                                                              & \emptycirc                                                     &           \\
\textbf{Foundational Framework}         & \fullcirc               & \emptycirc                                                   & \emptycirc                                                    & \halfcirc                                                        & \emptycirc                                              & \emptycirc                                                & \emptycirc                                                           & \emptycirc                                               & \emptycirc                                           & \emptycirc                                                             & \emptycirc                                                     &           \\
\textbf{Not Limited to Blockchain Identity} & \fullcirc           & \emptycirc                                                   & \halfcirc                                                     & \emptycirc                                                       & \emptycirc                                              & \emptycirc                                                & \fullcirc                                                            & \emptycirc                                               & \halfcirc                                            & \emptycirc                                                             & \halfcirc                                                      &           \\ \bottomrule
\end{tabular}
\end{adjustbox}
 \begin{tablenotes}
        \tiny
        \item \scalebox{0.75}{\fullcirc}\space included \scalebox{0.75}{\halfcirc}\space partially included \scalebox{0.75}{\emptycirc}\space omitted
\end{tablenotes}
\vspace{-0.75em}
\end{table}

Although not a survey, the closest work to ours is~\citet{AriesrfcsConcepts0207credentialfraudthreatmodel} who introduced a credential fraud threat model in a blockchain-based Aries RFC. 
 Our review of existing SSI surveys identified a lack of consensus on SSI components and their underlying design motivations. No existing work provides a comprehensive catalogue of SSI components, an in-depth analysis of trust assumptions, and an extensive collection of academic literature and industry implementations presented in this paper. Consequently, no existing work serves as a sufficient foundational framework for future research.
 
\subsection{Overview of SSI and Terminology} \label{sec:2_1_ssi_background}
\begin{figure}[ht]
\includegraphics[width=0.75\linewidth]{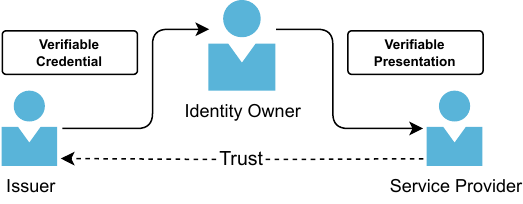}
\vspace{-1em}
\caption{SSI Credential Exchange}
\label{fig:ssi_simple}
\end{figure}

Self-sovereign identity (SSI) is founded on the issuance, storage, and presentation of identity credentials \cite{SovrinWhitepaper2018, SovrinGlossaryV3}. Three distinct roles are involved in SSI:

\noindent\textbf{Identity Owners} require identification and collect credentials that represent their identities. \\
\textbf{Issuers} are authorized entities that issue credentials to identity owners (government registries issuing driver's licenses). \\
\textbf{Service Providers} request credentials from identity owners to verify their identities. 

To facilitate further discussion, we clarify the definition of an identity. Aligned with the Sovrin Glossary \cite{SovrinGlossaryV3}, this work defines identity as a collection of \textit{attributes}, where each attribute represents a particular characteristic, such as name or age. Attributes are verified through attestations from issuers. These attestations are termed \textit{claims}. A collection of claims compiled by an issuer to describe an identity owner is a \textit{credential}. 

Credentials are stored in a personal wallet, typically on the identity owner's device. 
The exchange of credentials between parties is illustrated in Figure \ref{fig:ssi_simple}. Issuers share verifiable credentials with identity owners, who can then share verifiable presentations of these credentials with service providers. The additional term ``verifiable'' indicates that the credentials may be verified by the service providers; this concept is further discussed in Section \ref{sec:back:layer3}. Service providers rely on the credential issuer's authority without direct interaction. They infer the issuer's identity from the credential and use their existing knowledge of the issuer and their authority to determine whether to accept the credential (only a government authority can issue a driver's licence). This separation of credentials and isolation of the issuer from the service provider during credential presentations mitigate the concentration of issuer power that challenges the centralized identity models \cite{SovrinWhitepaper2018, allenPathSelfSovereignIdentity2016}.

This process of issuing, storing, and presenting identity credentials, known as the \textit{credential exchange}, is often portrayed as the entirety of SSI \cite{grunerAnalyzingComparingSecurity2023a, schardongSelfSovereignIdentitySystematic2022a}. While it represents a crucial component of SSI, it does not capture all of the components. For that, we will turn to the \textit{Trust Over IP Stack}.
  
\section{Trust Over IP and the Trustful Model for SSI} \label{sec:2_toip_background}
The Trust Over IP (ToIP) Stack introduced by the \textit{ToIP Foundation}\footnote{\url{https://trustoverip.org}} is a four-layer architectural framework designed to establish trust between users over the Internet and other digital networks. It is inspired by the TCP/IP stack, which standardized packet exchange and enabled the development of the Internet. The ToIP Stack, however, emphasizes human trust through establishing digital identities and secure communication. 

We outline the structure of SSI through the trust layers of the ToIP stack to analyze fundamental concepts and components of SSI in the context of forming trust.
Following the analysis, we introduce a minimal trust model, our baseline \textit{trustful model}.

\subsection{SSI through Trust Over IP Layers} \label{sec:2_toip}
\begin{figure}[ht]
\includegraphics[width=0.8\linewidth]{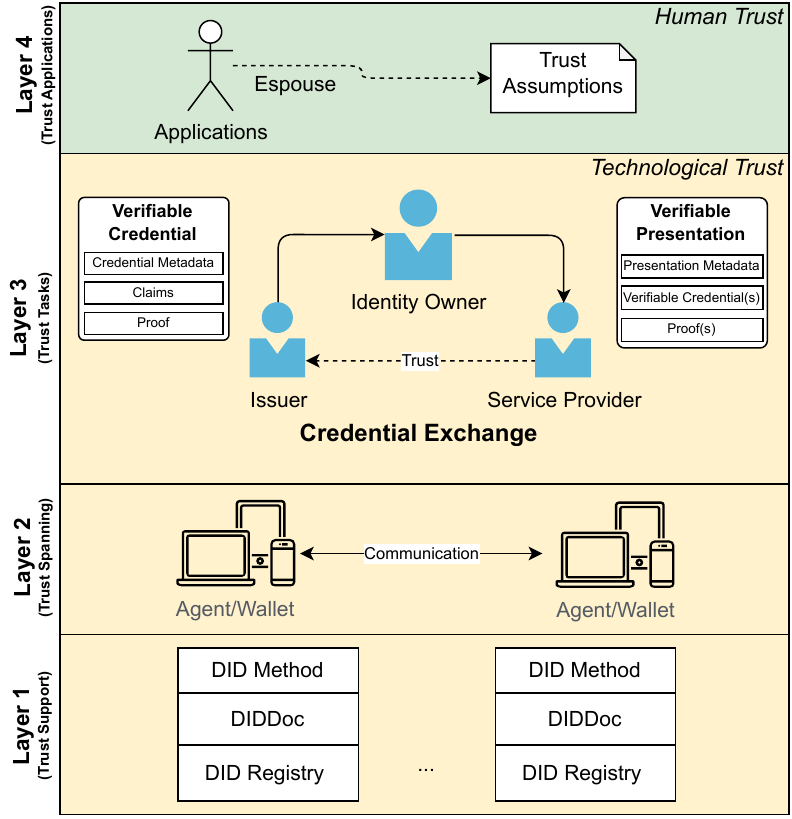}
\caption{SSI through Trust Over IP Stack}
\label{fig:ssi_toip}
\vspace{-1em}
\end{figure}

Our adaptation of ToIP layers tailored to SSI with the appropriate terminologies \cite{SovrinGlossaryV3} is shown in Figure \ref{fig:ssi_toip}. We walk through each layer, introduce the relevant components of SSI, and summarise implementation choices available for the components.

\subsubsection{Layer 1: Trust Support} \label{layer:1}
Layer \hyperref[layer:1]{1} provides a way to identify participants and verify the provenance of these identities through trust anchors. This layer provides a trusted foundation upon which the subsequent layers are constructed.

\paragraph{Identification}
Every party represents themselves with a globally unique self-generated identity handle \cite{ernstbergerSoKDataSovereignty}. Recipients communicating with the party use this identifier to derive public attributes (e.g., public key) \cite{DecentralizedIdentifiersDIDs}. Issuers create long-lived identifiers shared broadly, while other entities may create new identifiers for each communication. Two general implementation approaches to providing identifiers are decentralized identifiers (DIDs) or public keys.

DIDs are a URI of the format $did:method:identifier$. The DID \textit{method} specifies how the identifier is de-referenced. De-referencing a DID produces a DID Document (DIDDoc), which details the entity's public attributes. Every DIDDoc must include at least a public key for the subject \cite{DecentralizedIdentifiersDIDs}. Alternatively, a simple public key can serve as an identifier, eliminating the DID/DIDDoc abstraction\footnote{DID methods used for short-lived DIDs, where a trust anchor is not required, frequently require that the DID is the public key ($did:key$ \cite{DidKeyMethod},\space $did:peer$, \cite{PeerDIDMethod}). In these instances, DIDDoc resolution involves extracting the public key directly from the DID itself.}.

\paragraph{Trust Anchors}
Trust anchors affirm the provenance of identities and public attributes (DIDDocs). They are ``trusted components'' that serve as starting points for establishing trust across the other layers~\cite{IntroductionTrustIP2021}. For implementation, 
in distributed DID-based scenarios, verifiable data registries (VDRs, e.g., blockchain) often serve as trust anchors. However, public key infrastructure (PKI) may also be employed, notably when raw public keys are used instead of DIDs.

Recording a DIDDoc on a publicly accessible VDR, such as a blockchain, guarantees the record's immutability and auditability~\cite{EthrDIDResolver2023, SovrinDIDMethod}. The method associated with a VDR-backed DID specifies the location of the corresponding DIDDoc~\cite{bistarelliSurveyDecentralizedIdentifier}. For instance, \mbox{$did:ethr$} indicates the identifier is an Ethereum address that locates the DIDDoc~\cite{EthrDIDResolver2023}. Trust anchors are not always VDRs; \mbox{$did:web$} (and DID-less approaches) rely on traditional public key infrastructure to anchor identities and public attributes \cite{DidWebMethod}.

\vspace{0.85em}
\subsubsection{Layer 2: Trust Spanning} \label{layer:2}
Layer \hyperref[layer:2]{2} focuses solely on facilitating communication between agents and wallets of Layer \hyperref[layer:3]{3} participants. 
Agents, the software libraries managing credential requests, facilitate this communication and interact with wallets \cite{SovrinGlossaryV3}. For implementation, building upon DIDs (Layer \hyperref[layer:1]{1}), DIDComm \cite{DIDCommMessagingSpecification} establishes a suite of protocols for inter-entity messaging. DIDComm utilizes the public keys associated with DIDs to encrypt messages exchanged between entities.

\subsubsection{Layer 3: Trust Tasks} \label{sec:back:layer3} \label{layer:3}
Layer \hyperref[layer:3]{3} facilitates the credential exchange. It defines the standardized, interoperable credential formats (e.g., driver's licenses). It also specifies verification mechanisms (e.g., signatures), credential management protocols (collection, storage, presentation), and optional extensions to the core process.

\paragraph{Credential Formats}
Credentials adhere to a standardized format comprising credential metadata, claims, and proof. Credential metadata encompasses the issuer's identity (e.g., DID), credential type (e.g., driver's license), and issuance date, among other details~\cite{VerifiableCredentialsData}. 
Claims represent the identity attributes verified by the issuer (e.g., name, birthdate, address, licence no.). 
The proof ensures the credentials are tamper-proof for integrity and provenance. Presentations are temporary, ephemeral documents containing verifiable credentials that service providers request. Some works consider presentations synonymous with verifiable credentials~\cite{SovrinWhitepaper2018}. Regarding implementation, W3C verifiable credentials are the predominant data model, employing JSON-LD to define the credential structure~\cite{JSONLDJSONLinking, youngVerifiableCredentialsFlavorsa}. The AnonCreds JSON format links credential definitions to DIDs stored on VDR (Layer \hyperref[layer:1]{1})~\cite{AnonCredsSpecification2023}. 

\paragraph{Verification Mechanisms (Proofs)}

Credentials contain a verification mechanism as a \textit{proof} of the credential's authenticity. This proof may be provided through a signature, where the issuer signs the entire credential (metadata and claims) with their private key. Service providers verify the credential's authenticity by retrieving the issuer's public key from their DID. For implementation, ECDSA signatures are a common choice \cite{EllipticCurveDigital}. All surveyed credential profiles use signatures as a verification mechanism, although some do not present the signature to the service provider; instead, the identity owner presents a zero-knowledge proof of knowledge of a valid signature \cite{tessaroRevisitingBBSSignatures2023, camenischSignatureSchemeEfficient2003} (discussed further in Sections~\ref{sec:identity_owner:selective_disclosure}~and~\ref{sec:identity_owner:blinding}).

\paragraph{Protocols}
Protocols define the procedures controlling the exchange of credentials, encompassing how identity owners request credentials from issuers, how service providers request presentations from identity owners, and how identity owners present credentials to service providers. Many proposals implement various protocols. For instance, multiple related projects from OpenID address the issuing and presentation of verifiable credentials \cite{lodderstedtOpenIDVerifiableCredential2023, terbuOpenIDConnectVerifiable2022, yasudaSelfIssuedOpenIDProvider2023}. 

\paragraph{Extensions}
Extensions add additional SSI features to the credential exchange. Our discussion utilizes these extensions to introduce threats and analyze their risk and mitigations to derive various trust models (Section~\ref{sec:3_extensions}).  

\subsubsection{Layer 4: Trust Applications} \label{sec:2_toip_background:l4} \label{layer:4}
Layer \hyperref[layer:4]{4} depicts the identity applications ecosystem (e.g., government, health, banking). Each application is considered to have \textit{Trust Assumptions}. They define trusted and untrusted parties, interactions, and components. For instance, an application may distrust identity owners and assume they may present fraudulent credentials. 

\paragraph{Deriving Trust Requirements from Trust Assumptions}

Intuitively, a trust assumption implies a threat -- e.g., suggesting the credential presented to the service provider could be fraudulent.

In fact, we can derive that a trust assumption (or lack thereof) implies the application accepting the presence of a \textit{threat}. Threats pose a \textit{risk} to some participants and must be addressed through \textit{constraints}. That is, a Layer \hyperref[layer:4]{4} application that distrusts identity owners and assumes they may present fraudulent credentials mandates the implementation of tamper-proof credentials in Layer \hyperref[layer:3]{3}.

Based on this observation, in our analysis of trust requirements for trust models, we consider that the human trust assumptions made in Layer \hyperref[layer:4]{4} are stated as threats, which defines \textit{constraints} that need to be addressed by the components of Layers 1-3.

As a summary, Table \ref{tab:layers_cat} outlines common implementations or standardization initiatives for each SSI component discussed in this section. Trust assumptions in Layer \hyperref[layer:4]{4} determine the use of these components.

\begin{table}[h]
\vspace{-0.5em}
\caption{Common SSI Trust Over IP Layer Implementations}
\label{tab:layers_cat}
\begin{adjustbox}{width=0.47\textwidth,center}

\begin{tabular}{@{}lp{2.75cm}|p{8cm}@{}}
\toprule
\textbf{Layer \hyperref[layer:4]{4}}                  & \textbf{Application} & Various trust assumptions impacting choices Layer 1-3 \\ \midrule
\multirow{4}{*}{\textbf{Layer \hyperref[layer:3]{3}}} & \textbf{Format}                       & W3C VC~\cite{VerifiableCredentialsData} (JSON-LD~\cite{JSONLDJSONLinking}, JWT-VC~\cite{JWTVCPresentation}, SD-JWT-VC~\cite{terbuSDJWTbasedVerifiableCredentials2023}), AnonCred JSON~\cite{AnonCredsSpecification2023}, MDOC~\cite{businessWhereCanW3C2023}, ICOA DTC~\cite{GuidingCorePrinciples2020}, x.509, IRMA XML~\cite{IRMA}                             \\ \cmidrule(l){2-3} 
                                  & \textbf{Verification \mbox{Mechanism}}                  & ECDSA~\cite{EllipticCurveDigital}, RS256, CL~\cite{camenischSignatureSchemeEfficient2003}, BBS~\cite{tessaroRevisitingBBSSignatures2023}, BoundBBS~\cite{BoundBBSSignatures}, zkSNARK~\cite{nitulescuZkSNARKsGentleIntroduction}, GGM-Merkle~\cite{muktaBlockchainBasedVerifiableCredential2020b}                                                                                 \\ \cmidrule(l){2-3} 
                                  & \textbf{Protocol}                     & CHAPI~\cite{CHAPICredentialHandler}, WACI~\cite{WalletCredentialInteractions}, Aries Protocols~\cite{HyperledgerAriesrfcs2022}, OpenID4VCI~\cite{lodderstedtOpenIDVerifiableCredential2023}, OpenID4VP~\cite{terbuOpenIDConnectVerifiable2022}, SOIPv2~\cite{yasudaSelfIssuedOpenIDProvider2023}                                                    \\ \cmidrule(l){2-3} 
                                  & \textbf{Extensions}                   & \textit{Detailed in Section \ref{sec:3_extensions}} \\ \midrule
\textbf{Layer \hyperref[layer:2]{2}}                  & \textbf{Communication}                & DIDComm~\cite{DIDCommMessagingSpecification}                                                                                                        \\ \midrule
\multirow{2}{*}{\textbf{Layer \hyperref[layer:1]{1}}} & \textbf{Identification}               & DID (indy~\cite{IndyDIDMethod}, sov~\cite{SovrinDIDMethod}, ion~\cite{DecentralizedidentityIon2023}, ebsi~\cite{DIDMethodLegal2023}, key~\cite{DidKeyMethod}), Any Public Key (x.509~\cite{housleyInternet509Public1999})                             \\ \cmidrule(l){2-3} 
                                  & \textbf{Trust  Anchor}                & Blockchain (Indy DLT~\cite{IndySDK2023}, Bitcoin~\cite{DecentralizedidentityIon2023}, EBSI Trust Registries~\cite{IssuerTrustModel2023}, Ethereum~\cite{EthrDIDResolver2023}, Parity~\cite{BlockchainInfrastructureDecentralised}, Iroha~\cite{HyperledgerIroha}, Fabric~\cite{HyperledgerFabric})  IOTA~\cite{IOTA}, GNU Name System~\cite{schanzenbachGNUNameSystem2023}, Any PKI (CA, FIDO~\cite{FIDOAllianceOpen})                                         \\ \bottomrule
\end{tabular}

\end{adjustbox}
\vspace{-0.75em}
\end{table}
  
\subsection{The Trustful Model} \label{sec:3_semi_trusted_ssi}
Here, we introduce the first trust model: \textit{trustful} SSI. Trustful SSI encapsulates the fundamental principles of SSI, establishing the \textit{minimum responsibilities} for each component. As mentioned, the model is represented as a collection of threats accepted as posing risks under the application's trust assumptions. As a minimal model, it considers the threats relating to potential credential integrity and external actors, along with mitigation strategies. To clarify the design and responsibilities of the SSI components, we group the threats by components as a functional unit that implements a group of features.

\subsubsection{Identity Owner Identification} \label{sec:3:io_id}

\threatcmd{threat:semi:io:not_issued}
{
If identity owners can present credentials without a verification mechanism, they may present credentials (e.g., a driver's licence) that were never issued and do not represent their identity (e.g., the identity owner is not licensed to drive).
}{
A service provider may authorize an identity owner who should not have been authorized.
}{
Credentials (and the presentations that encompass them) must be verifiable. Credentials must be issued alongside some cryptographic proof that the service provider can use to verify the integrity of credentials.
}{
\textit{Identity owner identification} is implemented through the verifiable credentials in Layer \hyperref[layer:3]{3}. When the proof portion of verifiable credentials uses a signature to ensure the integrity of credentials the minimum requirement from the signature is three basic algorithms that provide \textit{key generation}, \textit{signing}, and \textit{verification} functions. These signatures must satisfy correctness and security \cite{EllipticCurveDigital} such that a service provider will always accept a valid signature by an issuer, and importantly, an adversary cannot forge a signature that passes the verification function but was not created by the issuer.  
}

\subsubsection{Issuer Identification} \label{sec:3:i_id}
\threatcmd{threat:semi:i:not_auth}{
An issuer may not have the authority to issue credentials of any type (e.g., may only issue state driver's licenses).
}{
A service provider may accept a credential issued by an entity that did not have the authority to attest to the identity claims. }{
The issuer of a credential must be identifiable from a verifiable presentation such that service providers can be assured that the claims they have been presented were issued by an issuer that they have some \textit{human trust} in (trust that a government entity may issue driver's licenses). 
}{
Layers \hyperref[layer:1]{1} and \hyperref[layer:3]{3} work together to address this constraint. Issuers embed and sign their DID in the Layer \hyperref[layer:3]{3} verifiable credentials. Service providers then use this DID to derive what entity issued the credential and verify their authority. The immutability of credentials (Threat~\ref{threat:semi:io:not_issued}) ensures that once issued, the issuer's DID can not be modified. The DID is then de-referenced to the publicly recorded DIDDoc to verify the identity and rights of the issuer.
}

\subsubsection{Non-Transferability} \label{sec:3:non_t}
\threatcmd{threat:semi:io:issued_to_other_io}{
The identity owner may present a correctly issued credential; however, it was not issued to them. Rather, it was collected through collusion with another identity owner or stolen from an unknowing identity owner (e.g., Alice presents Bob's driver's licence as if it was hers). 
}{
A service provider may authorize an identity owner by incorrectly believing claims that do not actually describe them.
}{
Verifiable credentials and their presentations must be tied to the identity owner such that they can not be transferred incorrectly to another identity owner.
}{
The identity owner's DID is included in the verifiable credential. The immutability (Section \ref{sec:3:io_id}) ensures that the DID cannot be modified after issuance. Ownership of the credential is provided by proving knowledge of the secret key associated with the public key of the DID (e.g., a Sigma protocol  \cite{damgaardSprotocols2002}). The issuer ensures that the identity owner knows the secret key when they embed the DID in the credential. Service providers perform the same verification when receiving presentations. For an adversary to present a verifiable credential as their own, they must have knowledge of the secret key associated with the embedded DID. 
}

\subsubsection{Protected Communication}
\threatcmd{threat:semi:outside:collect_cred_transit}{
An outside entity may intercept a credential in transit, either passively or actively. Passive interception would allow the outside entity to learn about the contents of the presentation, while active interception would allow the outside entity to modify the contents of the presentation. 
}{
For identity owners, their credentials in transit may unknowingly have their private identity information stolen or modified. 
}{
The credential exchange of Layer \hyperref[layer:2]{2} must be secured to protect communication between entities from outside entities. 
}{
The encryption in DIDComm protocols (Layer \hyperref[layer:2]{2}) ensures that no eavesdropper may learn the contents of verifiable presentations. Additionally, credentials cannot be modified by a third party if they are encrypted in transit. This includes the threats of replay \cite{syversonTaxonomyReplayAttacks1994a}, and man-in-the-middle attacks \cite{ganganReviewManintheMiddleAttacks2015} with protections such as token-binding in the Layer \hyperref[layer:3]{3} verifiable presentations~\cite{VerifiableCredentialData}.
}

\threatcmd{threat:semi:outside:did_resolution}{
An outside entity may interfere with DID resolution, resulting in an incorrect DIDDoc being de-referenced.
}{
A service provider may attempt to verify a credential using an incorrect DIDDoc of the issuer, resulting in either the authorization of an inauthentic or rejection of an authentic credential. 
}{
Layer \hyperref[layer:1]{1} must provide assurances in the provenance of resolved DIDDocs.
}{
When de-referencing a DID, service providers must be sure to use protected communication channels. Additionally, the registries publicly hosting the DIDDocs should provide provenance guarantees that the DIDDoc has not been modified. At the very least, these registries need to provide some level of assurance specified by the Layer \hyperref[layer:4]{4} application (e.g., blockchain-backed DID~\cite{SovrinDIDMethod, EthrDIDResolver2023}, certificate authority-backed DID~\cite{DidWebMethod}). 
}
 
\section{Extending Trust Requirements} \label{sec:3_extensions}
While adequate in many scenarios, trustful SSI may not fully address the intricacies and security demands of all digital identity applications. This is evident in government-issued digital identities, where the identity data may be highly sensitive. The trustful model's assumption of service provider trustworthiness overlooks potential misuse or insecure storage of identity data. 

This motivates more cautious approaches to SSI, where a broader set of trust requirements should be considered. For instance, since applications that follow the \textit{trustful} model do not consider the service providers a threat, identity owners may share their entire driver's licence credentials to prove their age. This is problematic as a dishonest service provider could collect the additional claims in a driver's licence for malicious purposes or collect them innocently but store them insecurely. 

Here, we broaden the trust requirements by introducing more threats perceived by issuers, identity holders and service providers.   

\subsection{For Issuers}\label{sec:prop_issuer}
The \textit{trustful model} did not include threats that impacted the issuers. 
Here, we introduce an extended set of threats for issuers.  

\subsubsection{Identity On-Boarding}
New identity owners may be brought into an SSI system through a process known as on-boarding. 

\threatcmd{threat:zero:i:identity_onboarding}{
Potential identity owners may attempt to mislead an issuer when being on-boarded such that they are issued credentials that do not represent their true identity (e.g., being issued a driver's licence credential with an incorrect name or driver's class).
}{
Issuers could be misled into issuing invalid credentials. 
}{
Credentials should only be issued if some assurances exist regarding the validity of the identity being attested. 
}{
Much of the issuer's responsibility in identifying a potential identity owner is offloaded outside technical constraints. Issuers are expected to perform some level of due diligence \cite{tobinSovrinWhatGoes2018}. Take, for example, a government-issued identity; it is reasonable to expect the government to have prior knowledge of the individual requesting a credential. Large portions of Sovrin's contributions in Layer \hyperref[layer:4]{4} have been creating frameworks and specifications that define the responsible behaviour of identity issuers to protect the system's integrity by having only responsible issuers prepare credentials \cite{SovrinWhitepaper2018}. This approach can be augmented with frequent auditings, such as what is done for certificate authorities in traditional public key infrastructure (PKI) \cite{archiveddocsSecuringPKIMonitoring2016}. Two optional extensions in Layer \hyperref[layer:3]{3} provide solutions to address this threat. 
\\
\indent\textit{\underline{Verifying Existing Credentials}; }
Taking the service provider role, issuers may request a credential that verifies common claims present in both an existing credential and the new one being pursued. If the claims align with the claims expected by the issuer, they issue the requested credential (e.g., a name claim matches another claim).
\\\indent 
To note, there is no indication within verifiable credentials that the issuer performed this verification. This approach must assume that the additional credential requested was not compromised through similar means. Finally, a potential identity owner may not have an existing SSI verifiable credential they can present, especially if they are a new user.
\\
\indent\textit{\underline{eID Derivation}; }
Deriving identity claims from outside SSI by bootstrapping SSI to existing identity schemes allows for low-friction identity on-boarding.
\citet{abrahamPrivacyPreservingEIDDerivation2020b} presented an SSI implementation that enabled deriving credentials from existing eID identity infrastructure. Identity owners created non-interactive zero-knowledge proofs \cite{nitulescuZkSNARKsGentleIntroduction} that their verifiable credentials were correctly derived from an existing eID system. 
}

\vspace{-1.55em}
\subsubsection{Issuance of Sensitive Claims}
\threatcmd{threat:zero:i:encrypted_claims}{
An identity owner may not have the right to view the contents of a credential issued to them. Such as when the credential is a confidential letter of recommendation regarding them. 
}{
An identity owner could collect sensitive claims, violating the privacy of the issuers. 
}{
Issuers should be able to issue credentials where claims are bound by a constraint that precludes parties -- including the identity owner -- from learning the contents of the claim. 
}{
A possible approach is \textit{\underline{Encrypted Claims}} where an issuer provides asymmetrically encrypted credentials with a secret key unknown to the identity owner. Identity owners present these credentials to service providers who know the secret key and can decrypt the claims. \citet{guajardoAnonymousCredentialSchemes2010a} introduced this in anonymous credentials. These constructions are limited as prior knowledge of the claim-accessing party is required but have found renewed interest through anonymous tokens \cite{chaseAnonymousTokensStronger2023, davidsonPrivacyPassBypassing2018}.
}
 
\subsection{For Identity Owners} \label{sec:prop_identity_owner}
In the \textit{trustful model}, the only threat to identity owners was the collection of credentials during transit. Here, we expand with the introduction of four new threats. 

\subsubsection{Selective Disclosure} \label{sec:identity_owner:selective_disclosure}
\textit{Data Minimalization}, collecting only necessary identity claims by the service providers,  is a fundamental principle of SSI \cite{allenPathSelfSovereignIdentity2016}. It is realized through the extension of \textit{selective disclosure} at Layer \hyperref[layer:3]{3}. 

\threatcmd{threat:zero:io:selective_disclosure}{
A service provider may collect more identity claims than required, either maliciously or without harmful intentions, such as a bar collecting an identity owner's name and address during age verification.
}{
Excessive collection of claims by service providers elevates the risks faced by identity owners when presenting credentials. Malicious service providers may exploit identity data for financial gain or profiling purposes \cite{cadwalladrRevealed50Million}, while even well-intentioned service providers inadvertently elevate identity loss risks with to their over-collection of identity data \cite{zouConcernNoAction2018}.
}{
Credential presentations must allow identity owners to present a subset of the claims in their credentials where service providers may not learn anything beyond what was presented. 
}{
Selective disclosure mechanisms can be achieved either through \textit{claim redaction} or \textit{predicates}.
\\
\indent\textit{\underline{Claim Redaction};} enables identity owners to remove claims from a verifiable credential when generating a presentation~\cite{zundelWhyVerifiableCredentials2021}. Service providers can verify the authenticity and integrity of the revealed claims without having access to the entire credential. Claim redaction describes an outcome where the service providers are only able to learn a subset of the claims that were initially included in a verifiable credential (e.g., a birthdate on a driver's license). Some approaches require that the identity owner only share the claims they want to disclose along with the accompanying proofs~\cite{duttoPostQuantumZeroKnowledgeVerifiable2022}; even though claims are not strictly being \textit{redacted}, the outcome remains the same. 
\\
\indent\textit{\underline{Predicates};} allows identity owners to provide a predicate proof (TRUE/FALSE) on a claim (e.g., their birthdate claim indicates they are over 18) without revealing the actual claim value.
The extent of the predicates available depends on the implementation chosen. Some implementations limit predicates to numerical ranges (e.g., age $>$ 18) \cite{camenischSignatureSchemeEfficient2003}, while others enable more complex assertions (e.g., the listed address is within some defined city boundary) \cite{leePrivacypreservingIdentityManagement2021a}.
\\\indent 
There are many proposals for implementing SSI selective disclosure categorized under three approaches: \textit{atomic} (signing each claim individually), \textit{hash records} (signing a hash of each claim rather than the claim itself, such that the service provider verifies the signature against the hashes of the claims and that the disclosed claim is the pre-image of the hash) \cite{terbuSDJWTbasedVerifiableCredentials2023}, and \textit{zero-knowledge proofs} (either as a custom proof designed for a specifically signature scheme~\cite{camenischSignatureSchemeEfficient2003, tessaroRevisitingBBSSignatures2023}, or a generic zero-knowledge proof on the claim and credential~\cite{leePrivacypreservingIdentityManagement2021a, nitulescuZkSNARKsGentleIntroduction}). While all of these approaches enable claim redaction, only \textit{zero-knowledge proofs} support genuine predicate selective disclosure. This functionality can be emulated by having the issuer sign a pre-computed predicate claim (i.e. $age\_over\_18 = TRUE$) during credential creation.
}

\subsubsection{Unlinkability} \label{sec:identity_owner:blinding}

\threatcmd{threat:zero:io:blinding_sd}{
Selectively disclosed claims provide limited privacy preservation if service providers can link presented claims to the same identity owner across consecutive presentations or through collusion with other service providers. Consider a venue using presentations to track patronage or a car rental service linking claims with information from an insurance agent without consent.
}{
If a presentation provided to a service provider contains an identifier unique to the identity owner, service providers can easily link an identity owner's actions \cite{chenAntiCollusionAnonymousCredentials2006a}. Service providers may then learn more about an identity owner's identity than what the identity owner consented to. 
}{
Verifiable presentations should not include any correlatable identifier the service provider can derive, such as identity owner DID or signature. 
}{
Single-show \cite{UProveCryptographicSpecification2023, baldimtsiAnonymousCredentialsLight2013} or \textit{$k$-show} \cite{layouniAnonymousKShowCredentials2007} credentials offer unlinkability only for a fixed (one or $k$) number of presentations of a credential. For effective SSI integration, credentials must enable identity owners to generate an unlimited number of unlinkable presentations from the same credential. Two unlinkability levels exist: \textit{complete} and \textit{limited}.
\\
\indent\textit{\underline{Complete Unlinkability};} To achieve complete unlinkability, where no presentations may ever be linked to the same credential, two transformations must be applied. The \textit{removal of correlating factors} addresses the vulnerability of a correlating DID identifier in every credential. Some works suggest managing the scope of DIDs, where the identity owner creates multiple DID pseudonyms, each used for a small subset of their credentials \cite{DidKeyMethod}. This still enables correlation for credentials issued to the same DID. Instead, verifiable credentials should not have DIDs (or other identifiers) embedded in them; DIDs remain only as a means for DIDComm communications and encryption (Layer \hyperref[layer:2]{2}) \cite{DIDCommMessagingSpecification} and issuer identification (Layer \hyperref[layer:1]{1}) \cite{tobinSovrinWhatGoes2018}. At times, this is the extent of unlinkability afforded (\citet{coelhoProposeFederatedLedger2018a}, \citet{takemiyaSoraIdentitySecure2018a}). 
\\\indent 
Still, the signature of the credential remains a correlating factor. With the \textit{transformation of signatures}, when the identity owner is preparing a credential for presentation, they generate a new \textit{proof} that they then provide in place of the signature from the issuer. This new proof is an unlinkable zero-knowledge proof of knowledge of the credential's signature. Here, anonymous credential signatures are used (BBS(+) \cite{camenischAnonymousAttestationUsing2016, tessaroRevisitingBBSSignatures2023}, CL \cite{camenischEfficientSystemNontransferable2001a}, PS \cite{pointchevalReassessingSecurityRandomizable2017}).
\\
\indent\textit{\underline{Limited Unlinkability};}
Complete unlinkability is not always desirable. In some situations, the service provider must be able to identify repeat presentations of identity owners, such as to prevent Sybil attacks \cite{maramCanDIDCanDoDecentralized2021}. If a service provider relies on SSI credentials for authentication into their service, such as through user accounts, then \textit{limited unlinkability} is required, allowing service providers to identify repeat presentations; otherwise, a single credential could be used to create many accounts. The limited unlinkability construction from \citet{zhangPASSOEfficientLightweight2021} included a unique identifier derived from a combination of the identity owner and the service provider's identifiers. Service providers may use this persistent identifier to identify repeat presentations but can not use it to correlate presentations made to other service providers. 
}

\vspace{1em}
\subsubsection{Key and Wallet Management} \label{sec:5_prop_identity_owner:key_and_wallet}

\threatcmd{threat:zero:io:protection}{
A malicious party could compromise and steal credentials or keys, or the identity owner may lose their keys.
}{
An identity owner who loses their keys or credentials (or has them compromised by a malicious party) then loses control over their personal identity information and its use.
}{
Credentials and keys must be protected in wallets, ensuring they are not vulnerable to being stolen or lost. 
}{
The risk of losing keys and, in turn, access to the credentials in the identity wallet are not concerns unique to SSI~\cite{maherCryptoBackupKey1996}. Potential solutions are: 
\\
\indent\textit{\underline{Trusted Third Party Key Recovery}}; \citet{soltaniPracticalKeyRecovery2019a} proposed a key recovery method using trusted third parties for SSI. 
The wallet of the identity owner is still in control of the user's credentials. As a precaution, third-party providers back up the keys in escrow. Approaches vary across implementations, but all either rely on some form of trusted third-party service~\cite{labordeUserCentricIdentityManagement2020a, linklaterDistributedKeyManagement2018a} or friend~\cite{naikUPortOpenSourceIdentity2020a, linklaterDistributedKeyManagement2018a}.
\\
\indent\textit{\underline{Multi-Factor Authentication (MFA)}; } 
With multi-factor authentication for credential presentations, the compromise of a wallet does not allow a malicious actor to present the compromised credentials as their own. \citet{zhangPASSOEfficientLightweight2021} implemented single sign-on using anonymous credentials following a structure compatible with Layer~\hyperref[layer:3]{3}. Service providers could require that identity owners present the same credentials from multiple devices. 
\\
\indent\textit{\underline{Multiple Devices}; } 
The same design used by \citet{zhangPASSOEfficientLightweight2021} for \textit{multi-factor authentication} also enables \textit{multiple devices}. Since credentials can be stored on any number of devices, identity owners could present their credentials from any registered device. 
}

\vspace{-1em}
\threatcmd{threat:zero:io:sovereignty}{
Identity owners cannot always be expected to follow best practices regarding safeguarding and maintaining access to their wallets. Without proper consideration, identity owners could rely on insecure wallet protection mechanisms (e.g., credentials stored on devices without access protection through passwords).
}{
Sacrificing credential security and sovereignty for usability places identity owners at risk of damaging their self-sovereignty (losing control of its use) or leakage of their credential data. 
}{
Identity owners must maintain control over their credentials and keys regardless of how they are stored \cite{allenPathSelfSovereignIdentity2016}. 
}{
A solution such as \textit{\underline{Cloud Wallets}; } places the entire responsibility of credential management with a cloud provider. Naively trusting a cloud provider with control over identity claims and their disclosure would violate the self-sovereignty of the identity owners (the cloud provider can dictate credential use). Furthermore, \citet{lagutinEnablingDecentralisedIdentifiers2019a} identified that constrained (e.g., embedded) devices might be unable to handle the resource overhead required by verifiable credentials. Their work ensured the constrained devices maintained control over their DIDs, but the credential processing was delegated to an authorization server. Notably, this implementation requires potentially unfounded trust in the authorization server and is only appropriate for SSI deployments in limited scopes. 
\\\indent 
To address these trust concerns, \citet{krennAttributeBasedCredentialsCloud2018}, later improved by \citet{habockBreakingFixingAnonymous2019a} presented anonymous credentials for the cloud. The proposed cloud wallets could never access the stored attributes in the clear, even when presenting the credentials. Their implementation utilized proxy re-encryption \cite{atenieseKeyPrivateProxyReencryption2009}, structure-preserving signatures \cite{fuchsbauerStructurePreservingSignaturesEquivalence2019a}, and zero-knowledge proofs of knowledge \cite{atenieseKeyPrivateProxyReencryption2009}. To provide transparency of a cloud wallet's actions, \citet{chaseCredentialTransparencySystem2022a} proposed a credential transparency system that allowed an identity owner to audit the actions of their cloud wallet.
}
 
\vspace{-4em}
\subsection{For Service Providers}  \label{sec:prop_service_provider}
Section \ref{sec:3_semi_trusted_ssi} introduced several threats service providers face; here, we expand on the threats faced by service providers and introduce new threats.

\subsubsection{Non-Transferability} \label{sec:sub:pop}

Previously, as a means to address non-transferability (Threat \ref{threat:semi:io:issued_to_other_io}), the concept of embedding the identity owner DID in the credential during issuance was introduced. However, this is not acceptable when \textit{unlinkability}  needs to be preserved. 
Here, we expand the mitigation approaches towards non-transferability through proving \textit{what you know} or \textit{what you own} with \textit{linkable} and \textit{unlinkable} variations of each. 

\indent\textit{\underline{What You Know};} 
DIDs embedded in a verifiable credential provide a suitable identifier to prove possession of a credential. This approach is the simplest and most often used \cite{muktaBlockchainBasedVerifiableCredential2020b}. This is \textit{visible non-transferability}, providing service providers with a linkable identifier in every credential presentation. 

A more robust solution is required to preserve \textit{unlinkability}; we classify these as \textit{hidden}. \citet{camenischSignatureSchemeEfficient2003}  presented protocols where the identity owners commit to a link secret when requesting a credential. The issuer verifies that the identity owner knows the committed link secret without learning the secret and signs the commitment. When identity owners present their credentials, they recommit to their link secret and share this pseudonym with the service provider. The service provider checks that the identity owner owns the committed secret. The identity owner also proves that the issuer signed a commitment on the same hidden secret. BBS signatures follow a similar principle \cite{BBSSignatureScheme}. Alternatively, the \textit{BoundBBS} proposal uses BLS signatures \cite{bonehShortSignaturesWeil} to bind credentials through knowledge of the secret BLS key without using the blind signing feature of BBS. Using BLS signatures removes the need to manage the representation of the link secrets and simplifies the BBS specification \cite{BoundBBSSignatures}.

\indent\textit{\underline{What You Have};} 
Physical ID cards often have photos of the holder to provide these same proof-of-possession assurances using biometrics. \citet{othmanHorcruxProtocolMethod2018} proposed the Horcrux Protocol, an SSI construction based on biometrics. This construction established DIDs linked to biometric factors,  tying credentials issued to a specific identity owner. However, this implementation did not provide the \textit{unlinkability} features and is a \textit{visible biometric} solution. 

As a \textit{hidden} implementation of biometric non-transferability, \citet{impagliazzoAnonymousCredentialsBiometricallyenforced2003a} introduced a scheme that tied anonymous credentials to biometric identifiers. Their construction enabled non-transferability without compromising the privacy of the identity owner. \citet{adamsAchievingNontransferabilityCredential2011a} and \citet{gerdesIncorporatingBiometricsVeiled2017a} presented improvements on this scheme. 

Credentials may also be tied to the identity owner's specific device (i.e., the device that hosts the credential wallet). These identifiers do not always have to be public/private key pairs: in the Internet of Things (IoT) context, \citet{niyaKYoTSelfsovereignIoT2020a} exploited the manufacturing variability of devices to produce a unique device identifier to embed in credentials.

\subsubsection{Credential Validity (Revocation)} \label{sec:sp:rev}
\threatcmd{threat:zero:sp:revocation}{
Credentials have a lifespan; identity owners can lose the privileges associated with a credential (e.g., having the right to drive revoked or a university identity card expiring after a student graduates). 
}{
A service provider who accepts a credential that the issuer has revoked is vulnerable to accepting identity owners who no longer have authority over the credentials they present. 
}{
Service providers must be able to determine if the credentials they are presented are still valid.
}{
We can broadly categorize the mitigating approaches as either \textit{expiring} or \textit{immediate}.
\\\indent
With \textit{\underline{Expiring}} credential validity, the issuer specifies a validity period. When a service provider is presented with a credential, they reject any credentials for which the validity period has passed. Credential validity can be renewed through the re-issuance of the credentials. Rather than requiring the whole credential to be re-issued, issuers can instead generate lightweight attestations to the validity of the credential that accompanies the presentation. Work by \citet{abrahamRevocableOfflineVerifiableSelfSovereign2020b} proposed a revocation scheme where the Layer \hyperref[layer:1]{1} network provides these attestations. \citet{camenischSolvingRevocationEfficient2010} provided a similar construction, where issuers regularly update and publish values that identity owners retrieve and use to re-validate their expired credentials. 
\\\indent
These short lifespans allow for efficient verification of credentials.
However, they are unable to provide immediate revocation, and a balance must be struck between having a short enough validity period that credentials can be revoked in an acceptable time frame while not being too short that issuers are over-burdened re-issuing or attesting to credential validity.
\\\indent
With \textit{\underline{Immediate}} revocation, the validity of a credential is altered with direct effect. Commonly, this is accomplished through a publicly available list of credential statuses (either listing the revoked or valid credentials). Issuers update this record if a credential must be revoked; service providers reference the list when verifying a credential. \citet{abrahamQualifiedEIDDerivation2018c} proposed using a revocation list utilizing the existing Layer \hyperref[layer:1]{1} distributed ledger.
\\\indent
Rather than maintaining a list of revoked or valid credentials, accumulators can collect the identifiers of all valid credentials into a single value that is published \cite{choVerifiableCredentialProof2021a}. When a credential is revoked, the accumulator is updated to no longer include the revoked credential. To preserve the \textit{unlinkability} requirement, the identity owner proves in zero-knowledge that their credential is registered in the dynamic accumulator \cite{camenischDynamicAccumulatorsApplication2002a}. 
\\\indent
Instead of a list or accumulator -- but still with direct effect -- \textit{reclaimID}, from \citet{schanzenbachReclaimIDSecureSelfSovereign2018b} used the online nature of their identity records to update the record's status. 
\\\indent
Compared to expiry credential validity, immediate revocation removes the burden of re-issuing credentials or re-publishing values for the issuers. In exchange, the identity infrastructure must accommodate the revocation or status list. 
}

\subsubsection{Identity Owner Identification}
\threatcmd{threat:zero:sp:identity_owner_self_attestation}{
Identity owners may not always be collecting credentials from trusted issuers. There may be situations where identity owners self-published claims of their identity. 
}{
Service providers are provided with no indication of the trustworthiness of an identity owner's self-published credentials.
}{
Presentations of self-published credentials should provide some confidence for service providers in the claims truth. 
}{
\textit{\underline{Reputation of Self Published Claims}} is an approach where reputation is built by collecting attestations to the validity of credentials in a web-of-trust approach \cite{stokkinkDeploymentBlockchainBasedSelfSovereign2018a}. Aside from variations across implementations and how reputation is collected and measured, implementations vary in their motivations as to why there is no trusted issuer. 
}

\vspace{-1.75em}
\threatcmd{threat:zero:sp:identity_owner_delegation}{
Identity owners may need to delegate the authority of their credentials to another identity owner in an exceptional circumstance (e.g., in healthcare). 
}{
Without delegation capabilities, identity owners who are in situations where they are unable to present their credentials are essentially left without their identity. 
}{
Credentials should enable their authority and claims to be shared with another identity owner when required. 
}{
For \textit{\underline{Delegation of Credentials}}, the identity owner takes on the role of an issuer, passing on the rights of their credential to a new identity owner \cite{muktaCredTrustCredentialBased2022}. Delegation can follow a chain, such as in the example of a hospital that issues a credential to a doctor who then issues the rights to their associates when they are on leave \cite{demuynckCredentialBasedSystemAnonymous2007a} or a patient who delegates a letter of their authority to a friend or family member \cite{muktaCredTrustCredentialBased2022}. \citet{chaseSignaturesKnowledge2006} (and \citet{demuynckCredentialBasedSystemAnonymous2007a}) introduced delegatable anonymous credentials as a further constraint that the delegating issuer should not be identifiable from the presentation of a delegated credential.
}

\vspace{-1.75em}
\threatcmd{threat:zero:sp:identity_owner_reputation}{
The ownership of a credential issued by a \textit{trustworthy} issuer does not indicate that the identity owner will behave appropriately. From Sections \ref{sec:identity_owner:selective_disclosure} (\textit{selective disclosure}) and \ref{sec:identity_owner:blinding} (\textit{unlinkability}), the service provider may not even be able to identify a misbehaving identity owner. Consider a case where SSI is used when renting a car; a driver's license credential, while proving identity and a right to drive, offers no guarantee of \mbox{safe driving}.
}{
Service providers may enter into a relationship with untrustworthy identity owners, and become unacceptably vulnerable.
}{
Service providers must be afforded some level of recourse or protection against identity owners who misbehave\footnote{The concept of misbehaving is situational to the identities use and the service provider's policies. }. 
}{
Using \textit{\underline{Blacklisting and Reputation}}, when \textit{unlinkability} is not considered, a service provider can easily blacklist a misbehaving identity owner from their service through their DID. With \textit{unlinkability}, this becomes a more complex problem. Additionally, one service provider may wish to report their experiences to other service providers while maintaining the identity owner's privacy. \citet{tsangPEREAPracticalTTPfree2008a} proposed a scheme that supported the blacklisting of anonymous credentials. \citet{nakanishiEfficientBlacklistableAnonymous2018a} expanded on this by allowing service providers to contribute to a reputation score of identity owners rather than a boolean blacklist. 
\\
\indent\textit{\underline{Anonymity Revocation}; } Another challenge introduced by \textit{unlinkability} is how service providers can take recourse (beyond just blacklisting from their service) against misbehaving identity owners. With roots in group signatures \cite{bonehShortGroupSignatures2004}, \textit{anonymity revocation} enables a service provider, supported by a mutually trusted third party between the identity owner and service provider, to revoke the anonymity provided by \textit{unlinkability}. \citet{camenischEfficientSystemNontransferable2001a} proposed that identity owners include some traceable identifier encrypted against the public key of a trusted revocation manager. The encrypted identifier included de-anonymization conditions to ensure the service provider could not tamper with the agreed-upon de-anonymization policy between the service provider and the identity owner. 
}

\subsubsection{Issuer Identification}
The authority to issue credentials is not enforced through SSI; this falls on the human trust and governance layer of SSI (Layer \hyperref[layer:4]{4}) \cite{SovrinWhitepaper2018}. As a brief expansion, some SSI proposals provide a means for the service provider to build trust in issuers that may not be \textit{traditionally} recognized via reputation. 

\textit{\underline{Issuer Reputation};}
From the same motivation of Threat \ref{threat:semi:i:not_auth}, \textit{issuer reputation} mechanisms enable service providers to learn some calculated reputation scores about an issuer. This reputation aims to measure the \textit{trustworthiness} of an issuer's attestations. \citet{bhattacharyaEnhancingSecurityPrivacy2020a} used the number of credentials issued by the issuer in a time window. Additionally, reputation may be measured against the specific claims of a credential rather than the issuer \cite{thomasEnhancingClaimBasedIdentity2009a}.

\subsubsection{Offline Verification} \label{sec:6_prop_service_provider:offline}
\threatcmd{threat:zero:sp:offline_verification}{
The identity owner's control of credentials in SSI lends itself well to offline operations where the identity owner or service provider may not have reliable Layer \hyperref[layer:1]{1} network access (e.g., verification of a credential in a remote region).
}{
Service providers may be denied from accepting (or identity owners presenting) valid credentials if their Layer \hyperref[layer:1]{1} network connection is interrupted or unavailable.
}{
The Layer \hyperref[layer:3]{3} credential exchange between identity owners and service providers must only rely on Layer \hyperref[layer:2]{2} peer-to-peer communications and not require access to the Layer \hyperref[layer:1]{1} network. 
}{
The self-contained proofs (signatures) in verifiable credentials permit offline verification. The peer-to-peer communications of Layer \hyperref[layer:2]{2} and pair-wise DIDs ensure neither the service provider nor the identity owner is required to reference the Layer~\hyperref[layer:1]{1} network during credential exchange. What is lost is the service provider's ability to de-reference issuer DIDs, find credential structures, and check revocation registries. Solutions to the DID and structure challenges are agnostic of implementation but introduce further constraints on the situations in which \textit{offline verification} could be performed. If the service provider is offline, they must know the credential structure they are requesting and have de-referenced the issuer's DID to their public key before being offline. 
\\\indent
In many cases, this is acceptable, especially if the service provider has intermittent access to the network to update their local cache. What poses a more significant challenge is Threat~\ref{threat:zero:sp:revocation} (\textit{credential validity}). Without immediate access to a revocation or status list, \textit{immediate} credential validity is incompatible. Credential status options are likely constrained to \textit{expiry} \cite{abrahamRevocableOfflineVerifiableSelfSovereign2020b}. Furthermore, \textit{key and wallet management} cannot rely on possibly inaccessible \textit{cloud wallets} (Threat \ref{threat:zero:io:sovereignty}). Some \textit{biometric} implementations for \textit{non-transferability} (Threat \ref{threat:semi:io:issued_to_other_io}) rely on server access for secure verification of credentials \cite{othmanHorcruxProtocolMethod2018}, making them incompatible with \textit{offline verification}. 
}
  
\section{The Trust Models} \label{sec:trust_models}
This section analyses the extended trust requirements, presented as threats and mitigations in Section~\ref{sec:3_extensions}, and introduces two additional models as alternatives to the \textit{trustful} model (Section~\ref{sec:3_semi_trusted_ssi}).

\subsection{Threat Dependencies}
\begin{figure}[ht]
\includegraphics[width=8cm]{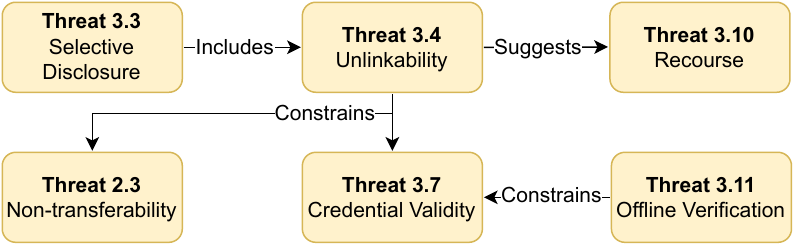}
\caption{SSI Threat Dependencies}
\label{fig:ssi-fpr}
\vspace{-1em}
\end{figure}

The discussions in Sections~\ref{sec:prop_issuer},~\ref{sec:prop_identity_owner}, and~\ref{sec:prop_service_provider} revealed implicit dependencies between threats as they emerged. Figure~\ref{fig:ssi-fpr} summarizes these dependencies. 

The privacy protection provided by selective disclosure is compromised if a malicious or colluding service provider may correlate presentations. This necessitates the inclusion of Threat~\ref{threat:zero:io:blinding_sd} (\textit{unlinkability}) if Threat~\ref{threat:zero:io:selective_disclosure} (\textit{selective disclosure}) is included in the trust assumptions.

Threat~\ref{threat:semi:io:issued_to_other_io} (\textit{non-transferability}) necessitates non-transferable credentials. Approaches that utilize an embedded identifier visible to service providers introduce a correlatable identifier that enables linking. In the presence of Threat~\ref{threat:zero:io:blinding_sd} (\textit{unlinkability}), \textit{hidden} non-transferability constructions must be employed to mitigate this risk.

Similarly, for Threat~\ref{threat:zero:sp:revocation} (\textit{credential validity}), approaches that employ credential validity lists must exclude correlatable identifiers from the lists and instead adopt an approach where identity owners prove their membership in an accumulator.

The anonymity potentially provided by \textit{unlinkability} (Threat~\ref{threat:zero:io:blinding_sd}) may place the service providers at risk, suggesting Threat~\ref{threat:zero:sp:identity_owner_reputation} (\textit{blacklisting, reputation, and anonymity revocation}).

The incorporation of Threat~\ref{threat:zero:sp:offline_verification} (\textit{offline verification}) imposes limitations on Threat~\ref{threat:zero:sp:revocation} (\textit{credential validity}), restricting implementations that require Layer \hyperref[layer:1]{1} access (immediate via lists).

The consideration of these dependencies is relevant to discussions involving alternative models and the range of trust assumptions within each model. Opting to address one threat may require incorporating another related threat mitigation.

\subsection{The Trust Models of SSI} \label{sec:models:trust}
\begin{table}[ht]
\renewcommand{\arraystretch}{0.5} \caption{Trust Models and Recommended Threats}
\label{tab:models}
\begin{adjustbox}{width=.47\textwidth,center}
\begin{tabular}{@{}lcccc@{}}
\toprule
\textbf{Extension / Feature}                                        & \textbf{Threat} & \textbf{Trustful} & \textbf{Intermediate Trust} & \textbf{Zero-Trust} \\ \midrule
\multirow{4}{*}{\textbf{Identity Owner Identification}} & \textbf{\ref{threat:semi:io:not_issued}}    & \fullcirc         & \fullcirc                   & \fullcirc           \\
                                                        & \textbf{\ref{threat:zero:sp:identity_owner_self_attestation}}    & \emptycirc        & \emptycirc                  & \fullcirc           \\
                                                        & \textbf{\ref{threat:zero:sp:identity_owner_delegation}}    & \emptycirc        & \emptycirc                  & \emptycirc          \\ 
                                                        & \textbf{\ref{threat:zero:sp:identity_owner_reputation}}    & \emptycirc        & \halfcirc                   & \halfcirc           \\ \midrule
\textbf{Issuer Identification}                          & \textbf{\ref{threat:semi:i:not_auth}}    & \fullcirc         & \fullcirc                   & \scalebox{1.5}{$\otimes$} \\ \midrule
\textbf{Non-Transferability}                            & \textbf{\ref{threat:semi:io:issued_to_other_io}}    & \fullcirc         & \fullcirc                   & \fullcirc           \\ \midrule
\multirow{2}{*}{\textbf{Protected Communication}}       & \textbf{\ref{threat:semi:outside:collect_cred_transit}}    & \fullcirc         & \fullcirc                   & \fullcirc           \\
                                                        & \textbf{\ref{threat:semi:outside:did_resolution}}    & \fullcirc         & \fullcirc                   & \fullcirc           \\ \midrule
\textbf{Identity On-Boarding}                           & \textbf{\ref{threat:zero:i:identity_onboarding}}    & \emptycirc        & \emptycirc                  & \scalebox{1.5}{$\otimes$} \\ \midrule
\textbf{Issuance of Sensitive Claims}                   & \textbf{\ref{threat:zero:i:encrypted_claims}}    & \emptycirc        & \emptycirc                  & \emptycirc          \\ \midrule
\textbf{Selective Disclosure}                           & \textbf{\ref{threat:zero:io:selective_disclosure}}    & \emptycirc        & \fullcirc                   & \fullcirc           \\ \midrule
\textbf{Unlinkability}                                  & \textbf{\ref{threat:zero:io:blinding_sd}}    & \emptycirc        & \fullcirc                   & \fullcirc           \\ \midrule
\multirow{2}{*}{\textbf{Key and Wallet Management}}     & \textbf{\ref{threat:zero:io:protection}}    & \emptycirc        & \halfcirc                   & \halfcirc           \\
                                                        & \textbf{\ref{threat:zero:io:sovereignty}}    & \emptycirc        & \halfcirc                   & \halfcirc           \\ \midrule
\textbf{Credential Validity (Revocation)}                                     & \textbf{\ref{threat:zero:sp:revocation}}    & \halfcirc         & \fullcirc                   & \fullcirc           \\ \midrule
\textbf{Offline Verification}                           & \textbf{\ref{threat:zero:sp:offline_verification}}    & \emptycirc        & \emptycirc                  & \emptycirc          \\ \bottomrule
\end{tabular}
\end{adjustbox}
 \begin{tablenotes}
        \tiny
        \item \scalebox{0.75}{\fullcirc}\space required \scalebox{0.75}{\halfcirc}\space likely \scalebox{0.75}{\emptycirc}\space optional extension $\otimes$\space forbidden
    \end{tablenotes}
\vspace{-1.5em}
\end{table}
 
The design of SSI components is influenced by underlying trust assumptions, often implicitly embedded in academic literature and industry implementations. We identify and categorize these implicit trust assumptions into three distinct trust models for SSI, representing the overarching approaches to trust. These trust models define the scope of what is untrusted in the Layer \hyperref[layer:4]{4} application; effectively, the models identify the threats users face. The three models represent varying levels of trust, with each subsequent model assuming less trust, accepting less risk, and subsequently capturing more threats than the previous. We further explore the implications of these assumptions and provide recommendations on what threats are captured by each model in the \textit{Implications on Threat Mitigation} sections. Table~\ref{tab:models} summarizes the recommended threats captured by each trust model, providing a reference of the remaining discussions in this section.

\subsubsection{Trustful Model}
\paragraph{General Concept}
The \textit{trustful} model (from Section~\ref{sec:3_semi_trusted_ssi}) assumes as few threats as possible. Implementation approaches that follow this trust model often \textit{do not consider the sensitivity of the identity data to be high}. 
This model assumes \textit{complete trust in issuers} regarding credential issuance and indicates \textit{no trust concerns about service providers' behaviour}.

\paragraph{Implications on Threat Mitigation}

As the most trusting model, the captured threats must provide a functional identity system. Credentials should be verifiable by service providers (Threat~\ref{threat:semi:io:not_issued} \textit{identity owner identification}). Service providers should be able to ascertain the issuer of a credential (Threat~\ref{threat:semi:i:not_auth} \textit{issuer identification}) and the ownership of credentials (Threat~\ref{threat:semi:io:issued_to_other_io} \textit{non-transferability}). Additionally, the Layer \hyperref[layer:2]{2} communication channels employed and the Layer \hyperref[layer:1]{1} identification should provide assurances of security (Threats \ref{threat:semi:outside:collect_cred_transit} and \ref{threat:semi:outside:did_resolution} \textit{protected communication}).

Implementing an SSI system under this model necessitates a rudimentary Layer~\hyperref[layer:1]{1} DID infrastructure for issuer identification, Layer~\hyperref[layer:2]{2} protocols like DIDComm \cite{DIDCommMessagingSpecification} to safeguard communications, and a credential format coupled with a \textit{simple} verification mechanism such as an ECDSA signature \cite{EllipticCurveDigital}. 

\subsubsection{Intermediate Trust Model}
\paragraph{General Concept}

The \textit{intermediate trust model} extends upon the \textit{trustful model} with a lower risk tolerance, making it more suitable for applications handling sensitive data. Service providers are no longer trusted, and all entities adopt a more cautious approach to potential threats. Implementations adhering to the \textit{intermediate trust model} often target large-scale identity systems, such as those used for government-issued credentials, as opposed to internal company authentication tools.

\paragraph{Implications on Threat Mitigation}

To safeguard identity owners from the now untrusted service providers, Threat~\ref{threat:zero:io:selective_disclosure} (\textit{selective disclosure}) should be considered. The lower risk tolerance also indicates that Threat~\ref{threat:zero:sp:revocation} (\textit{credential validity}) should be included. Following the dependencies outlined in Figure~\ref{fig:ssi-fpr}, Threat~\ref{threat:zero:io:blinding_sd} (\textit{unlinkability}) would also need to be included. The implementation of \textit{credential validity} (Threat~\ref{threat:zero:sp:revocation}) and Threat~\ref{threat:semi:io:issued_to_other_io} (\textit{non-transferability}) must be restricted to their \textit{unlinkable/hidden} variants, and Threat~\ref{threat:zero:sp:identity_owner_reputation} (\textit{blacklisting, reputation, and anonymity revocation}) may be considered viable.

Implementations do not necessitate modifications to Layer \hyperref[layer:1]{1}~or~\hyperref[layer:2]{2}. Layer~\hyperref[layer:3]{3} would require the selection of a signature scheme that supports \textit{selective disclosure} and \textit{unlinkability}. Furthermore, the format and protocols may need to be extended to accommodate the additional extensions.

\subsubsection{Zero-Trust Model}
\paragraph{General Concept}
The \textit{zero trust model} challenges the notion of a predefined set of trusted issuers with absolute authority to verify identities. Under this assumption, any form of central authority is considered a threat that demands mitigation. These systems embody a fully decentralized identity system.

\paragraph{Implications on Threat Mitigation}

To eliminate the concept of trusted issuers, Threat~\ref{threat:semi:i:not_auth} (\textit{issuer identification}) is removed and prohibited from the model. Threat~\ref{threat:zero:i:identity_onboarding} (\textit{identity on-boarding}) is also incompatible. Threat~\ref{threat:zero:sp:identity_owner_self_attestation} (\textit{self-attestation of claims}) is introduced to augment \textit{identity owner identification}. Beyond these changes, the low-risk tolerance implies the same trust assumptions as the \textit{intermediate trust model}.

In fully \textit{trustless} systems, the centralized authority of traditional public key infrastructure (PKI) is no longer appropriate and, if present, is replaced with decentralized trust anchors, typically in the form of a verifiable data registry (VDR) such as a blockchain. 

The absence of established issuers necessitates reconsidering the Layer \hyperref[layer:3]{3} credential exchange structure. Credentials and their corresponding proof mechanisms must support alternative assurance mechanisms, such as web-of-trust and self-attestation of claims. 
 
\section{The State of SSI} \label{sec:ssi_state}
This section examines the current landscape of self-sovereign identity (SSI) from the perspective of the components and threat assumptions discussed earlier. Both academic literature and industry implementations are evaluated. The component selections and the approaches adopted for the extensions are presented. Columns were omitted in instances where no information was available. 

Table \ref{table:ssi} presents a compilation of academic contributions gathered through the literature review outlined in Appendix \ref{sec:app:methodology}. The double horizontal line separates works that present a complete SSI construction (top) from those that only introduce a variation of an extension (bottom). Table \ref{tab:ssi_real} lists industry implementations. These implementations are commonly referred to as credential profiles~\cite{CredentialComparisonMatrix}. In both tables, ($\--$) indicates that the component or extension was not considered in the implementation. Table \ref{tab:ssi_real_cat} further categorizes the approaches employed for the extensions in the credential profiles.

\begin{table}[ht]
    \caption{Credential Profile Extension Categorizations}
    \label{tab:ssi_real_cat}
    \begin{adjustbox}{width=0.47\textwidth,center}
    
    \begin{tabular}{@{}lll|p{6cm}@{}}
    \toprule
    \multirow{3}{*}{\textbf{Selective Disclosure}} & \textbf{None}                           &                   & RS256 \cite{MicrosoftEntraVerified}                                                                                 \\ \cmidrule(l){2-4} 
    & \textbf{Redact}                         &                   & ECDSA\textsuperscript{*} \cite{EllipticCurveDigital}, BBS(+) \cite{tessaroRevisitingBBSSignatures2023}, BoundBBS \cite{BoundBBSSignatures}, GGM-Merkle \cite{muktaBlockchainBasedVerifiableCredential2020b}                                   \\ \cmidrule(l){2-4} 
    & \textbf{Predicates}                     &                   & CL \cite{camenischSignatureSchemeEfficient2003}, ZKP (zkSNARK \cite{nitulescuZkSNARKsGentleIntroduction})                                                                     \\ \midrule
\multirow{3}{*}{\textbf{Unlinkability}}        & \textbf{None}                           &                   & ECDSA, RS256, GGM-Merkle, BLS                                                         \\ \cmidrule(l){2-4} 
    & \textbf{Limited}                        &                   & \--                                                                                   \\ \cmidrule(l){2-4} 
    & \textbf{Complete}                       &                   & CL, BBS(+), BoundBBS, ZKP (zkSNARK)                                                      \\ \midrule
\multirow{4}{*}{\textbf{Non-Transferability}}  & \multirow{2}{*}{\textbf{What you know}} & \textbf{Linkable} & DID Key, JWT \cite{auth0.comJWTIO}, Any Key                                                                 \\ \cmidrule(l){3-4} 
    &                                         & \textbf{Hidden}   & Link Secret, BLS Key (BoundBBS)                                                       \\ \cmidrule(l){2-4} 
    & \multirow{2}{*}{\textbf{What you have}} & \textbf{Linkable} & Biometric or Device                                                                   \\ \cmidrule(l){3-4} 
    &                                         & \textbf{Hidden}   & \--                                                                                   \\ \midrule
\multirow{3}{*}{\textbf{Revocation}}           & \textbf{None}                           &                   & \--                                                                                   \\ \cmidrule(l){2-4} 
    & \textbf{Expiry}                         &                   & Epoch Expiry                                                                          \\ \cmidrule(l){2-4} 
    & \textbf{Immediate}                      &                   & AnonCreds Revocation \cite{AnonCredsRevocation2023}, W3C Status List 2021 \cite{BitstringStatusList}, EBSI Status List \cite{OverviewEBSIRevocation2023}, JWT and CWT Status List \cite{lookerJWTCWTStatus2023} \\ \bottomrule
\end{tabular}

\end{adjustbox}
 \begin{tablenotes}
        \tiny 
        \item \begin{minipage}[t]{0.1cm}
            \textbf{*}
        \end{minipage} 
        \begin{minipage}[t]{8cm} 
            When used with \textit{SD-JWT-VC} redaction is possible \cite{terbuSDJWTbasedVerifiableCredentials2023}. \textit{ECDSA} does not offer selective disclosure with the other credential formats in Table \ref{tab:ssi_real}.
        \end{minipage}
\end{tablenotes}
\end{table}
 \begin{table*}[ht]
\caption{Academic Contributions Comparison}
\label{table:ssi}
\begin{adjustbox}{width=1\textwidth,center}
\begin{tabular}{@{}lllllllllllllllp{1.5cm}@{}}
\toprule
                                                                                          & \textbf{}             & \textbf{}               & \textbf{}          & \textbf{}                         & \textbf{\rotDown{Selective Disclosure}} & \textbf{\rotDown{Unlinkability}} & \textbf{\rotDown{Non-Transferability}} & \textbf{\rotDown{Credential Validity}} & \textbf{\rotDown{Issuer Identification}} & \textbf{\rotDown{Key and Wallet Management}} & \textbf{\rotDown{Identity On-Boarding}} & \textbf{\rotDown{Issuance of Sensitive Claims}} & \textbf{\rotDown{Identity Owner Identification}} & \textbf{\rotDown{Offline Verification}} &  \\
                                                                                          & \textbf{Trust Anchor} & \textbf{Identification} & \textbf{Signature} & \textbf{Trust Assumption}         & \textbf{}                               & \textbf{}                        & \textbf{}                              & \textbf{}                              & \textbf{}                                & \textbf{}                                    & \textbf{}                               & \textbf{}                                       & \textbf{}                                        & \textbf{}                               &  \\ \midrule
\multicolumn{1}{l|}{\textbf{\citet{coelhoProposeFederatedLedger2018a}}}                   & Blockchain            & Public Key              & \--                & \multicolumn{1}{l|}{Intermediate} & $\times$                                & \--                              & Know                                   & Either                                 & \--                                      & \--                                          & \--                                     & \--                                             & \--                                              & \--                                     &  \\
\multicolumn{1}{l|}{\textbf{\citet{niuSelfsovereignIdentityManagement2021a}}}             & Ethereum              & Public Key              & \--                & \multicolumn{1}{l|}{Intermediate} & $\times$                                & \--                              & Know                                   & Immediate                              & \--                                      & TTP                                          & \--                                     & \--                                             & \--                                              & $\times$                                &  \\
\multicolumn{1}{l|}{\textbf{\citet{stokkinkTrulySelfSovereignIdentity2021a}}}             & Any PKI               & Public Key              & ZKP                & \multicolumn{1}{l|}{Intermediate} & $\times$                                & \checked                         & Have                                   & Either                                 & \--                                      & \--                                          & \--                                     & \--                                             & \--                                              & $\times$                                &  \\
\multicolumn{1}{l|}{\textbf{\citet{labordeUserCentricIdentityManagement2020a}}}           & FIDO PKI              & FIDO                & \--                & \multicolumn{1}{l|}{Trustful}     & \--                                     & \--                              & Know                                   & \--                                    & \--                                      & FIDO                                         & \--                                     & \--                                             & \--                                              & $\times$                                &  \\
\multicolumn{1}{l|}{\textbf{\citet{borseAnonymitySecureIdentity2019a}}}                   & Ethereum              & Public Key              & ZKP                & \multicolumn{1}{l|}{Intermediate} & \--                                     & \checked                         & Know                                   & \--                                    & \--                                      & \--                                          & \--                                     & \--                                             & \--                                              & $\times$                                &  \\
\multicolumn{1}{l|}{\textbf{\citet{muktaBlockchainBasedVerifiableCredential2020b}}}       & Parity                & DID                     & GGM-Merkle         & \multicolumn{1}{l|}{Intermediate} & $\times$                                & \--                              & Know                                   & \--                                    & \--                                      & \--                                          & \--                                     & \--                                             & \--                                              & \--                                     &  \\
\multicolumn{1}{l|}{\textbf{\citet{lueckingDecentralizedIdentityTrust2020a}}}             & IOTA                  & DID                     & \--                & \multicolumn{1}{l|}{Zero-Trust}   & \--                                     & \--                              & Know                                   & \--                                    & \--                                      & \--                                          & \--                                     & \--                                             & Self-Publish                                     & $\times$                                &  \\
\multicolumn{1}{l|}{\textbf{\citet{stokkinkDeploymentBlockchainBasedSelfSovereign2018a}}} & Blockchain            & Public Key              & ZKP                & \multicolumn{1}{l|}{Zero-Trust}   & $\leq$                                  & \--                              & Know                                   & Expiry                                 & \--                                      & MFA                                          & \--                                     & \--                                             & Self-Publish                                     & \--                                     &  \\
\multicolumn{1}{l|}{\textbf{\citet{alsayedkassemDNSIdMBlockchainIdentity2019}}}           & Ethereum              & Public Key              & \--                & \multicolumn{1}{l|}{Intermediate} & $\times$                                & \--                              & Know                                   & \--                                    & \--                                      & \--                                          & \--                                     & \--                                             & \--                                              & $\times$                                &  \\
\multicolumn{1}{l|}{\textbf{\citet{lagutinEnablingDecentralisedIdentifiers2019a}}}        & Indy                  & DID                     & \--                & \multicolumn{1}{l|}{Trustful}     & \--                                     & \--                              & Know                                   & \--                                    & \--                                      & \--                                          & \--                                     & \--                                             & Delegation                                       & $\times$                                &  \\
\multicolumn{1}{l|}{\textbf{\citet{abrahamPrivacyPreservingEIDDerivation2020b}}}          & Blockchain            & DID                     & BLS                & \multicolumn{1}{l|}{Intermediate} & $\times$                                & \--                              & Know                                   & Immediate                              & \--                                      & \--                                          & eID                                     & \--                                             & \--                                              & \--                                     &  \\
\multicolumn{1}{l|}{\textbf{\citet{leePrivacypreservingIdentityManagement2021a}}}         & Blockchain            & DID                     & zkSNARK            & \multicolumn{1}{l|}{Intermediate} & $\leq$                                  & \checked                         & Know                                   & \--                                    & \--                                      & \--                                          & \--                                     & \--                                             & \--                                              & \--                                     &  \\
\multicolumn{1}{l|}{\textbf{\citet{hamerPrivateDigitalIdentity}}}                         & Indy                  & DID                     & \--                & \multicolumn{1}{l|}{Intermediate} & \--                                     & \checked                         & Have                                   & Immediate                              & \--                                      & \--                                          & \--                                     & \--                                             & \--                                              & \--                                     &  \\
\multicolumn{1}{l|}{\textbf{\citet{abrahamQualifiedEIDDerivation2018c}}}                  & Indy                  & DID                     & \--                & \multicolumn{1}{l|}{Trustful}     & \--                                     & \--                              & Know                                   & Immediate                              & \--                                      & \--                                          & eID                                     & \--                                             & \--                                              & \--                                     &  \\
\multicolumn{1}{l|}{\textbf{\citet{schanzenbachReclaimIDSecureSelfSovereign2018b}}}       & GNU Name System       & Public Key              & \--                & \multicolumn{1}{l|}{Zero-Trust}   & $\times$                                & \--                              & Know                                   & Expiry                                 & \--                                      & \--                                          & \--                                     & \--                                             & Self-Publish                                     & $\times$                                &  \\
\multicolumn{1}{l|}{\textbf{\citet{abrahamRevocableOfflineVerifiableSelfSovereign2020b}}} & Blockchain            & DID                     & BLS                & \multicolumn{1}{l|}{Intermediate} & $\leq$                                  & \--                              & Know                                   & Expiry                                 & \--                                      & \--                                          & \--                                     & \--                                             & \--                                              & \checked                                &  \\
\multicolumn{1}{l|}{\textbf{\citet{takemiyaSoraIdentitySecure2018a}}}                     & Iroha                 & DID                     & \--                & \multicolumn{1}{l|}{Intermediate} & $\times$                                & \--                              & Know                                   & \--                                    & \--                                      & \--                                          & \--                                     & \--                                             & Self-Publish                                     & $\times$                                &  \\
\multicolumn{1}{l|}{\textbf{\citet{choVerifiableCredentialProof2021a}}}                   & Indy                  & DID                     & CL                 & \multicolumn{1}{l|}{Intermediate} & \--                                     & \checked                         & Know                                   & Immediate                              & \--                                      & \--                                          & \--                                     & \--                                             & \--                                              & \--                                     &  \\ \midrule \midrule
\multicolumn{1}{l|}{\textbf{\citet{grunerQuantifiableTrustModel2018a}}}                   & Blockchain            & \--                     & \--                & \multicolumn{1}{l|}{Zero-Trust}   & \--                                     & \--                              & \--                                    & \--                                    & \--                                      & \--                                          & \--                                     & \--                                             & Self-Publish                                     & \--                                     &  \\
\multicolumn{1}{l|}{\textbf{\citet{lauingerAPoAAnonymousProof2021a}}}                     & Indy                  & DID                     & \--                & \multicolumn{1}{l|}{Intermediate} & \--                                     & \--                              & \--                                    & \--                                    & Authority                                & \--                                          & \--                                     & \--                                             & \--                                              & \--                                     &  \\
\multicolumn{1}{l|}{\textbf{\citet{chakravartyBlockchainenhancedIdentitiesSecure2018a}}}  & Blockchain            & DID                     & \--                & \multicolumn{1}{l|}{Trustful}     & \--                                     & \--                              & \--                                    & \--                                    & \--                                      & \--                                          & \--                                     & \--                                             & \--                                              & \--                                     &  \\
\multicolumn{1}{l|}{\textbf{\citet{muktaCredTrustCredentialBased2022}}}                   & Fabric                & DID                     & \--                & \multicolumn{1}{l|}{Intermediate} & \--                                     & \--                              & \--                                    & \--                                    & \--                                      & \--                                          & \--                                     & \--                                             & Delegation                                       & \--                                     &  \\
\multicolumn{1}{l|}{\textbf{\citet{bhattacharyaEnhancingSecurityPrivacy2020a}}}           & Indy                  & DID                     & \--                & \multicolumn{1}{l|}{Intermediate} & \--                                     & \--                              & Know                                   & \--                                    & Reputation                               & \--                                          & \--                                     & \--                                             & \--                                              & \--                                     &  \\
\multicolumn{1}{l|}{\textbf{\citet{zolotavkinImprovingUnlinkabilityAttributebased2022}}}  & \--                   & \--                     & \--                & \multicolumn{1}{l|}{Intermediate} & $\leq$                                  & \checked                         & \--                                    & \--                                    & \--                                      & \--                                          & \--                                     & \--                                             & \--                                              & \--                                     &  \\
\multicolumn{1}{l|}{\textbf{\citet{niyaKYoTSelfsovereignIoT2020a}}}                       & Ethereum              & Public Key              & \--                & \multicolumn{1}{l|}{Trustful}     & \--                                     & \--                              & Have                                   & \--                                    & \--                                      & \--                                          & \--                                     & \--                                             & \--                                              & \--                                     &  \\
\multicolumn{1}{l|}{\textbf{\citet{soltaniPracticalKeyRecovery2019a}}}                    & Indy                  & DID                     & \--                & \multicolumn{1}{l|}{Trustful}     & \--                                     & \--                              & \--                                    & \--                                    & \--                                      & TTP                                          & \--                                     & \--                                             & \--                                              & \--                                     &  \\
\multicolumn{1}{l|}{\textbf{\citet{othmanHorcruxProtocolMethod2018}}}                     & Blockchain            & DID                     & \--                & \multicolumn{1}{l|}{Trustful}     & \--                                     & \--                              & Have                                   & \--                                    & \--                                      & \--                                          & \--                                     & \--                                             & \--                                              & $\times$                                &  \\
\multicolumn{1}{l|}{\textbf{\citet{linklaterDistributedKeyManagement2018a}}}              & Any PKI               & Public Key              & \--                & \multicolumn{1}{l|}{Trustful}     & \--                                     & \--                              & \--                                    & \--                                    & \--                                      & TTP + MD                                     & \--                                     & \--                                             & Delegation                                       & \checked                                &  \\
\multicolumn{1}{l|}{\textbf{\citet{grunerUsingProbabilisticAttribute2019a}}}              & Blockchain            & DID                     & \--                & \multicolumn{1}{l|}{Zero-Trust}   & \--                                     & \--                              & \--                                    & \--                                    & \--                                      & \--                                          & \--                                     & \--                                             & Self-Publish                                     & \--                                     &  \\ \bottomrule
\end{tabular}
\end{adjustbox}
 \begin{tablenotes}
        \tiny
        \item \textbf{Selective Disclosure: } $\leq$ \space = predicate $\times$\space = claim redaction \space\textbf{Unlinkability:} \checked\space = complete \space \textbf{Offline Verification: } \checked\space = supported $\times$\space = prohibited \textbf{ZKP: } Zero-Knowledge Proof
        \item \textbf{Key and Wallet Management:} MD = multi-device TTP = trusted third-party MFA = multi-factor authentication FIDO = key management offloaded to FIDO components \cite{FIDOAllianceOpen} 
\end{tablenotes}
\end{table*}
 
\begin{table*}[t]
\caption{Credential Profiles Comparison}
\label{tab:ssi_real}
\begin{adjustbox}{width=1\textwidth,center}
\begin{tabular}{@{}llllllllp{1cm}llllp{0cm}@{}}
\toprule
                                                             & \multicolumn{1}{c}{\textbf{}}             & \multicolumn{2}{c}{\textbf{Identification}}                                   & \multicolumn{2}{c}{\textbf{Credential Exchange}}                             & \multicolumn{1}{c}{\textbf{}}                 & \textbf{\down{S.D.}} & \textbf{\down{Unlink.}} & \textbf{\down{Non-Transferability}} & \textbf{\down{Credential Validity}} & \textbf{\down{O.V.}} & \textbf{\down{K \& W Mgmt.}} & \textbf{} \\ 
\textbf{Credential}                                          & \multicolumn{1}{l}{\textbf{Trust Anchor}} & \multicolumn{1}{l}{\textbf{Issuer}} & \multicolumn{1}{l}{\textbf{Identity Owner}} & \multicolumn{1}{l}{\textbf{Format}} & \multicolumn{1}{l}{\textbf{Signature}} & \multicolumn{1}{l}{\textbf{Trust Assumption}} & \textbf{}                               & \textbf{}                        & \textbf{}                              & \textbf{}                              & \textbf{}                               & \textbf{}                              & \textbf{} \\ \midrule
\multicolumn{1}{l|}{\textbf{Hyperledger   AnonCreds \cite{AnonCredsSpecification2023}}}        & Indy DLT                                  & did:indy                            & Link Secret                                 & AnonCred JSON                       & CL                                     & \multicolumn{1}{l|}{Intermediate}             & $\leq$                                  & \checked                         & Link Secret                            & AnonCreds Revocation                   & $\sim$                                  & MD / TTP / CD                          &           \\
\multicolumn{1}{l|}{\textbf{JSON-LD +   BoundBBS(+) \cite{Jsonldsignaturesbbs2023}}}        & CA PKI                                    & did:web                             & BLS Public Key                              & JSON-LD                             & BoundBBS(+)                            & \multicolumn{1}{l|}{Intermediate}             & $\times$                                & \checked                         & BLS Key                                & W3C Status List 2021                   & $\sim$                                  & \--                                    &           \\
\multicolumn{1}{l|}{\textbf{JSON-LD +   BBS(+) \cite{Jsonldsignaturesbbs2023}}}             & CA PKI                                    & did:web                             & did:key                                     & JSON-LD                             & BBS(+)                                 & \multicolumn{1}{l|}{Trustful}                 & $\times$                                & \--                              & DID Key                                & W3C Status List 2021                   & $\sim$                                  & \--                                    &           \\
\multicolumn{1}{l|}{\textbf{JWT-VC \cite{JWTVCPresentation}}}                         & Bitcoin                                   & did:ion                             & did:ion                                     & JWT-VC                              & ECDSA                                  & \multicolumn{1}{l|}{Trustful}                 & \--                                     & \--                              & DID Key                                & W3C Status List 2021                   & $\sim$                                  & \--                                    &           \\
\multicolumn{1}{l|}{\textbf{JWT-VC \cite{JWTVCPresentation}}}                         & Bitcoin + CA PKI                          & did:web                             & did:ion                                     & JWT-VC                              & ECDSA                                  & \multicolumn{1}{l|}{Trustful}                 & \--                                     & \--                              & DID Key                                & W3C Status List 2021                   & $\sim$                                  & \--                                    &           \\
\multicolumn{1}{l|}{\textbf{NGI   Atlantic for OpenID4VCs \cite{NextGenerationSSI}}}  & CA PKI                                    & did:web                             & did:key                                     & JWT-VC                              & ECDSA                                  & \multicolumn{1}{l|}{Trustful}                 & \--                                     & \--                              & DID Key                                & Epoch Expiry                           & $\sim$                                  & \--                                    &           \\
\multicolumn{1}{l|}{\textbf{NGI   Atlantic for OpenID4VCs \cite{NextGenerationSSI}}}  & Any PKI                                   & did:key                             & did:key                                     & JWT-VC                              & ECDSA                                  & \multicolumn{1}{l|}{Trustful}                 & \--                                     & \--                              & DID Key                                & Epoch Expiry                           & $\sim$                                  & \--                                    &           \\
\multicolumn{1}{l|}{\textbf{Dutch   Decentralized Identity \cite{DutchDecentralizedIdentity2023}}} & CA PKI                                    & did:web                             & did:jwt                                     & JWT-VC                              & ECDSA                                  & \multicolumn{1}{l|}{Trustful}                 & \--                                     & \--                              & jwt                                    & W3C Status List 2021                   & $\sim$                                  & \--                                    &           \\
\multicolumn{1}{l|}{\textbf{Microsoft   Entra  did:web \cite{MicrosoftEntraVerified}}}     & CA PKI                                    & did:web                             & did:web                                     & JWT-VC                              & RS256                                  & \multicolumn{1}{l|}{Trustful}                 & \--                                     & \--                              & jwt                                    & W3C Status List 2021                   & $\sim$                                  & \--                                    &           \\
\multicolumn{1}{l|}{\textbf{Microsoft   Entra  did:ion \cite{MicrosoftEntraVerified}}}     & Bitcoin                                   & did:ion                             & did:ion                                     & JWT-VC                              & RS256                                  & \multicolumn{1}{l|}{Trustful}                 & \--                                     & \--                              & jwt                                    & W3C Status List 2021                   & $\sim$                                  & \--                                    &           \\
\multicolumn{1}{l|}{\textbf{Selective   Disclosure JWT \cite{terbuSDJWTbasedVerifiableCredentials2023}}}     & Any PKI                                   & Any Public Key                      & Any Public Key                              & SD-JWT-VC                           & ECDSA                                  & \multicolumn{1}{l|}{Trustful}                 & $\times$                                & \--                              & jwt                                    & W3C Status List 2021                   & $\sim$                                  & \--                                    &           \\
\multicolumn{1}{l|}{\textbf{ESSIF \cite{ESSIFLab}}}                          & EBSI Trust Registries                     & did:ebsi                            & did:ebsi                                    & SD-JWT-VC                           & ECDSA                                  & \multicolumn{1}{l|}{Trustful}                 & $\times$                                & \--                              & DID Key                                & EBSI Status List                       & $\sim$                                  & \--                                    &           \\
\multicolumn{1}{l|}{\textbf{OpenID4VC   High Assurance \cite{yasudaOpenID4VCHighAssurance2023}}}     & Any PKI                                   & Any Public Key                      & Any Public Key                              & SD-JWT-VC                           & ECDSA                                  & \multicolumn{1}{l|}{Trustful}                 & $\times$                                & \--                              & Any Key                                & JWT and CWT Status List                & $\sim$                                  & \--                                    &           \\
\multicolumn{1}{l|}{\textbf{ISO mDL   (ISO/IEC 18013-5) \cite{businessWhereCanW3C2023}}}    & x.509  + CA PKI                           & X.509 Public Key                    & COSE Public Key                             & MDOC                                & ECDSA                                  & \multicolumn{1}{l|}{Trustful}                 & $\times$                                & \--                              & Biometric and/or Device                & Epoch Expiry                           & \checked                                & MD                                     &           \\
\multicolumn{1}{l|}{\textbf{ICAO DTC \cite{GuidingCorePrinciples2020}}}                       & Any PKI                                   & Any Public Key                      & Any Public Key                              & ICOA DTC                            & ECDSA                                  & \multicolumn{1}{l|}{Trustful}                 & \--                                     & \--                              & Biometric and/or Passport              & Cerfticate Revocation List             & $\sim$                                  & \--                                    &           \\
\multicolumn{1}{l|}{\textbf{x.509 \cite{housleyInternet509Public1999}}}                          & Any PKI                                   & Any Public Key                      & Any Public Key                              & x.509                               & ECDSA                                  & \multicolumn{1}{l|}{Trustful}                 & \--                                     & \--                              & jwt                                    & Cerfticate Revocation List             & $\sim$                                  & \--                                    &           \\
\multicolumn{1}{l|}{\textbf{IRMA   (Yivi Wallet) \cite{WhatIRMAIRMAa}}}           & Any PKI                                   & Any Public Key                      & Link Secret                                 & IRMA XML                            & CL                                     & \multicolumn{1}{l|}{Intermediate}             & $\leq$                                  & \checked                         & Link Secret                            & CL Revocation                          & \checked                                & \--                                    &           \\ \bottomrule
\end{tabular}
\end{adjustbox}
 \begin{tablenotes}
        \tiny
        \item \textbf{S.D.: } Selective Disclosure \space \textbf{Unlink.: } Unlinkability \space \textbf{O.V.: } Offline Verification \space \textbf{K \& W Mgmt.: } Key and Wallet Management
        \item \textbf{Selective Disclosure: } $\leq$ \space = predicate $\times$\space = claim redaction \space\textbf{Unlinkability:} \checked\space = complete \--\space = none \space\textbf{Key and Wallet Management:} MD = multi-device TTP = trusted third-party CW = cloud wallets \space\textbf{Offline Verification: } \checked\space = supported $\sim$\space = possible
        \item \textbf{Aggregation of Signatures, Identity On-Boarding, Issuance of Sensitive Claims, Identity Owner Identification} excluded from table as no credential profile considered them. 
\end{tablenotes}
\end{table*}
 
Several notable distinctions between academic and industry SSI implementations underscore the divergence between the two domains. Most prominently, the choice of trust anchors differs significantly. Academic constructions predominantly rely on blockchains, such as the identity-specific Indy platform \cite{HyperledgerIndyHyperledger}, as trust anchors. This extensive use of blockchains has led some to erroneously categorize SSI as a blockchain identity. However, this is inaccurate, as the majority of credential profiles in Table \ref{tab:ssi_real} utilize \textit{traditional} public key infrastructure. This fundamental difference stems from varying trust assumptions and differing risk priorities. None of the industry implementations adheres to the \textit{zero-trust} model, which disallows the existence of trusted authorities. Instead, all industry implementations permit some level of trusted entity, whereas many academic contributions do not and necessitate \textit{zero-trust}.

Still, there are commonalities across domains. Both academic and industry approaches often share fundamental goals, such as agreeing on broad trust models for privacy protection (intermediate and zero-trust models) or aiming for a practical digital identity solution (trustful model). Despite these shared objectives, differing risk perspectives lead to variations in component choices within the overarching framework. Notably, these implementation differences do not indicate significant changes in the framework or architecture. Instead, adapting to address potential threats often requires simply swapping specific components within the established structure.

The trust assumptions listed for each contribution of Tables \ref{table:ssi} and \ref{tab:ssi_real} reflect the authors' implicit threat assumptions, aligning with the models presented in Section \ref{sec:trust_models}. Comparing the implemented constraints or mitigations through the selected components and extensions against the recommended threats for each model category reveals that few works fully address the threats implied by the \textit{intermediate} or \textit{zero-trust} models.

These discrepancies do not imply any fault in the authors' implementations; however, they do emphasize the need for explicit threat definitions and careful consideration of the requisite extensions and components. Ultimately, achieving mutual agreement on a framework to capture SSI as well as a catalogue of threats and mitigations available can foster a more consistent approach to SSI across all domains.

\section{Open Work} \label{sec:open_work}
\paragraph{\textbf{Post-Quantum Resistance;}}

Existing signature schemes employed in SSI systems, as outlined in Table \ref{table:ssi} and \ref{tab:ssi_real}, lack quantum-safe resilience. This vulnerability extends beyond Layer \hyperref[layer:3]{3} signatures to encompass the cryptographic underpinnings of Layers~\hyperref[layer:1]{1} and \hyperref[layer:2]{2}, potentially jeopardizing the entire cryptographic trust architecture depicted in Figure~\ref{fig:ssi_toip}. While post-quantum secure solutions have emerged for public-key management in Layers~\hyperref[layer:1]{1}~and~\hyperref[layer:2]{2}, Layer~\hyperref[layer:3]{3} solutions remain limited. Though \citet{jeudyLatticeSignatureEfficient2022} proposed a quantum-secure anonymous credential scheme suitable for Layer~\hyperref[layer:3]{3} signatures, \citet{duttoPostQuantumZeroKnowledgeVerifiable2022} highlighted the continued challenge of achieving efficient selective disclosure (predicates or redaction). 

\paragraph{\textbf{Interoperability of Credentials;}}
Without broad interoperability of SSI credentials, identity owners will again be siloed into identity systems that have chosen compatible constructions. Interoperability must be considered for each of the layers. Layer \hyperref[layer:1]{1} constructions that use DIDs conforming to the standards of \cite{DecentralizedIdentifiersDIDs} are, by design, interoperable. Any entity that needs to dereference a DID can follow the specification of the defined method and collect the standardized DIDDoc. Credential and presentation incompatibilities represent the most significant obstacle to interoperability. The protocols, signatures, credential format and definitions, claims canonicalization, and encoding methods must all be compatible.

Verifiable credentials based on JSON-LD, such as the W3C VC Data Model \cite{VerifiableCredentialsData}, claim compatibility with JSON credentials \cite{youngMisinformationStopsHere2023}, but this has been disputed \cite{KristinaYasudaLinkedIn}. AnonCreds employs a proprietary credential format \cite{AnonCredsSpecification2023}, deviating from W3C verifiable credential standards. The ISO 18013-5 mobile Driver's License proposal also adheres to a proprietary format \cite{businessWhereCanW3C2023}.

SSI interoperability remains an ongoing challenge, and to date, there is no definitive consensus on what this should look like. The disparity of implementations in Tables~\ref{table:ssi}~and~\ref{tab:ssi_real} serves as evidence.

\paragraph{\textbf{Key Management;}} 
Section \ref{sec:5_prop_identity_owner:key_and_wallet} introduced the concept of multi-factor authentication (via presentations). The proposal by \citet{zhangPASSOEfficientLightweight2021} required that identity owners present their credentials multiple times from distinct devices. This is inefficient as all that is required is that identity owners are protected against a single on-device wallet compromise. 

In addition, the protection of keys used by issuers when signing credentials through key rotation has been a source of contention within the standards \cite{NeedClarifyRevocation, IssuerKeyRotation}. The \textit{online} nature of resolved DIDDocs provides many methods for issuers to rotate or revoke their signing keys. With rotation, a credential issued under a key pre-rotation is considered valid if issued when the key was valid. Revocation of a signing key indicates that the key has been compromised, and existing credentials signed under the key should not be accepted. This is further complicated as assuming an issuer is malicious and holds a compromised key, the issue date embedded in the credential cannot be trusted. \textit{AnonCreds} relies on the Layer 1 node attesting to the issuance date, further complicating issuance and verification and introducing further challenges to \textit{offline verification} (Section~\ref{sec:6_prop_service_provider:offline}). Protecting these keys and their lifecycles is critical to the resilience of SSI, and yet, there is a gap in addressing this challenge. 

\paragraph{\textbf{Transparency of Service Providers;}} 
Service providers establish trust in issuers through either pre-existing trust relationships (using DIDs to link the trusted issuer with the credential signer) or credential reputation (Threat~\ref{threat:zero:sp:identity_owner_reputation}). They build trust in identity owners through the Layer 3 credential exchange. What is missing is the \textit{transparency of service providers}. Identity owners must hope that the service provider they interact with is honest. Often, it is up to the identity owner to evaluate the risk of disclosure.

Some discussions of digital identities, particularly those from government authorities, propose that the issuing entity accredit service providers (see Australia's proposed digital identity law \cite{AustraliaDigitalID}). However, this approach risks further siloing identities and hindering interoperability. The current state-of-the-art for protecting identity owners from malicious service providers offers two options: limiting service providers to a pre-approved list of accredited entities or relying on \textit{selective disclosure} and \textit{unlinkability} to safeguard identity owners from unfettered service providers. The effectiveness of these measures remains uncertain, and there is a strong likelihood that they may not be sufficient. Consider a scenario where a service provider must collect a claim that uniquely identifies the identity owner (i.e., a student number id) because the service provider does not trust the identity owner, then any attempt at \textit{unlinkability} mitigation's is destined to fail. To address this, the discussion of Threat~\ref{threat:zero:sp:identity_owner_reputation} introduced some mitigations (\textit{blacklisting, reputation, and anonymity revocation}) that may enable a service provider to \textit{find} trust in an \textit{unlinkable} identity owner. Yet, this balance between the competing motivations of trust and risk priorities remains under-explored, with no clear solution \mbox{to appease both parties.}

\paragraph{\textbf{Unlinkability Remains Ambiguous;}} 

Section~\ref{sec:identity_owner:blinding} delves into the complexities surrounding \textit{unlinkability} in SSI. Correlating \textit{unlinkability} and \textit{anonymity} is a typical response when presented with \textit{unlinkability}. Anonymous credentials -- one of the core technologies in many SSI constructions -- even bear the name. However, this does not adequately communicate the nuances. \textit{Anonymity} encompasses a broader scope than \textit{unlinkability}. To provide \textit{anonymity} would require that the SSI construction prevent tracking of identity owners through means beyond signatures in the presentations. Correlatable identifiers can be extracted from network protocols. TOR \cite{TorSecondGenerationOnion} may be able to address this in an \textit{online} credential exchange; peer-to-peer credential exchanges proposed on top of layers such as NFC, Bluetooth, and QR Codes also face this challenge and will require consideration if total \textit{anonymity} is required. 

None of this matters, if the disclosed claims place identity owners in small identity, sets such that they could be profiled regardless. It remains disputed if applications should take on the burden of unlinkability if \textit{anonymity} cannot be provided (evident by the number of credential profiles in Table~\ref{tab:ssi_real} that provide \textit{selective disclosure} without \textit{unlinkability}). At a minimum, future contributions must clarify their position on \textit{unlinkability}, communicating what level of \textit{unlinkability} their construction provides to identity owners. 
 
\section{Conclusion} \label{sec:takeaways}

As digital identities become unavoidable, Self-Sovereign Identity (SSI) offers a promising direction for individual control and privacy. Through a systematic analysis of SSI's components and examining trust assumptions and threats posed by various actors, we developed three distinct trust models. These models capture the diverse levels of trust found across different SSI implementations and literature. We highlight the multifaceted nature of trust in SSI and how various stakeholders perceive and approach trust differently. Our comprehensive catalogue of SSI components and trust-related design requirements serves as a valuable resource for practitioners and researchers.
\\
\indent \textit{For practitioners. }
Practitioners tasked with designing an SSI deployment are provided with a framework to build their construction (Section~\ref{sec:2_toip}). The trust models from Section~\ref{sec:trust_models} define a broad categorization and reference of the assumed risks. The catalogue of components, design requirements, and existing implementations in Sections~\ref{sec:3_semi_trusted_ssi},~\ref{sec:3_extensions},~and~\ref{sec:ssi_state} may be referenced to select the appropriate mitigations based on the specific threats identified in the chosen trust model.
\\
\indent \textit{For researchers. }
Researchers can use this same framework and catalogue of components to locate their work in the field of SSI. Additionally, we support the justification of further work through our trust models and threats and identify open, unaddressed work.

\begin{acks}
The work has been supported by the Cyber Security Research Centre Limited whose activities are partially funded by the Australian Government’s Cooperative Research Centres Programme.
This research has been supported by an Australian Government Research Training Program (RTP) Scholarship.
\end{acks}

\bibliographystyle{ACM-Reference-Format}
\bibliography{refs}

\appendix
\section{Methodology} \label{sec:app:methodology}
We started our systematization with the seed article \textit{The Path to Self-Sovereign Identity} by Allen \cite{allenPathSelfSovereignIdentity2016}. References to this article were collected through ResearchRabbit\footnote{\url{https://researchrabbitapp.com/}}. These articles were filtered and included if they met the following requirements:

\begin{enumerate}
    \item The work proposed an SSI implementation.
    \item or; the work proposed a feature for an existing SSI implementation.
    \item or; the work met the prior two requirements for some close adjacent field.
\end{enumerate}

We also defined the following hard exclusion requirements:
\begin{enumerate}
    \item The work did not propose any SSI features or implementations.
    \item or; the work discussed a use case of SSI, that is, an implementation of existing SSI tooling for a specific scenario.
    \item or; the work was purely focused on cryptographic protocols without practical application.
    \item or; on point 3 from the inclusion requirements, the adjacent implementation did not have a clear analog to SSI (federations, PKI,  authentication without attributes).
\end{enumerate}

We collected all of the selected papers' references and citations from the same database and repeated the filtering process. We continued these steps until the filtering converged.

This systematization resulted in two broad categories of SSI work: academic literature and industry implementations. We included extra consideration for the latter with additional sources of industry credential profiles\footnote{\url{https://github.com/WebOfTrustInfo/rwot11-the-hague/blob/master/draft-documents/credential-profile-comparison.md}} as they have received more attention in development and deployment in recent years than purely academic discussions. Losing their contributions would not provide a complete image of the state-of-the-art.
 
\end{document}